\documentclass[draft,tightenlines,nofootinbib,preprint,aps,eqsecnum,amsmath,amssymb]{revtex4}

\newcommand{\beq}{\begin{equation}}
\newcommand{\eeq}{\end{equation}}
\newcommand{\bea}{\begin{eqnarray}}
\newcommand{\eea}{\end{eqnarray}}

\begin{document}

\title{
Explaining Leibniz-equivalence as difference of non-inertial
appearances: dis-solution of the Hole Argument \\ and physical
individuation of point-events}

\medskip

\author{Luca Lusanna}

\affiliation{ Sezione INFN di Firenze\\ Polo Scientifico\\ Via Sansone 1\\
50019 Sesto Fiorentino (FI), Italy\\ Phone: 0039-055-4572334\\
FAX: 0039-055-4572364\\ E-mail: lusanna@fi.infn.it}

\author{Massimo Pauri}

\affiliation{
Dipartimento di Fisica - Sezione Teorica\\ Universita' di Parma\\
Parco Area Scienze 7/A\\ 43100 Parma, Italy; \\
and Sezione INFN di Milano Bicocca - Gruppo Collegato di Parma.\\
Phone: 0039-0521-905219\\ FAX: 0039-0521-905223\\
E-mail:
pauri@pr.infn.it\\
\ \\
\today\\
 }

\begin{abstract}
\ \\
"The last remnant of physical objectivity of space-time" is
disclosed in the case of a continuous family of spatially
non-compact models of general relativity (GR). The {\it physical
individuation} of point-events is furnished by the autonomous
degrees of freedom of the gravitational field, (viz, the {\it Dirac
observables}) which represent - as it were - the {\it ontic} part of
the metric field. The physical role of the {\it epistemic} part
(viz. the {\it gauge} variables) is likewise clarified as embodying
the unavoidable non-inertial aspects of GR. At the end the
philosophical import of the {\it Hole Argument} is substantially
weakened and in fact the Argument itself dis-solved, while a
specific four-dimensional {\it holistic and structuralist} view of
space-time (called {\it point-structuralism}) emerges, including
elements common to the tradition of both {\it substantivalism} and
{\it relationism}. The observables of our models undergo real {\it
temporal change}: this gives new evidence to the fact that
statements like the {\it frozen-time} character of evolution, as
other ontological claims about GR, are {\it model dependent}.

\bigskip
\bigskip

{\bf KEYWORDS}:

\medskip

- Hole Argument

- Leibniz equivalence

- Structuralism

- Dirac observables

- Gauge variables

- Non-inertial frames

\end{abstract}

\maketitle

\vfill\eject

\section{Introduction}

The fact that the requirement of general covariance might involve a
threat to the physical objectivity of the points of space-time as
represented by the theory of gravitation was becoming clear to
Einstein even before the theory he was trying to construct was
completed. It was during the years 1913-1915 that the threat took
form with the famous Hole Argument ({\it Lochbetrachtung})
(Einstein, 1914) \footnote{For a beautiful historical critique see
Norton 1987, 1992, 1993.}. In the literature about classical field
theories space-time points are usually taken to play the role of
individuals, but it is often implicit that they can be distinguished
only by the physical fields they carry. Yet, the Hole Argument
apparently forbids precisely this kind of individuation, for it
entails that different - so-called diffeomorphic - {\it models} of
general relativity (GR) be taken as {\it physically equivalent},
under the menace of {\it indeterminism} for the theory. Since, on
the other hand, the Argument is a direct consequence of the {\it
general covariance} of GR, this conflict eventually led Einstein to
state (our {\it emphasis}):
\begin{quotation}
\footnotesize \noindent That this {\it requirement of general
covariance}, which {\it takes away from space and time the last
remnant of physical objectivity}, is a natural one, will be seen
from the following reflexion... (Einstein, 1916, p.117).
\end{quotation}

Although Einstein quickly bypassed the seeming cogency of the Hole
Argument against the implementation of general covariance on the
purely instrumentalist grounds of the so-called {\it
Point-Coincidence Argument}\footnote{The assertion that {\it "the
reality of the world-occurrence (in opposition to that dependent on
the choice of reference system) subsists in space-time
coincidences."}}, the issue remained in the background of the theory
until the Hole Argument received new life in recent years with a
seminal paper by John Stachel (1980). This paper, followed seven
years later by Earman and Norton's philosophical argument against
the so-called space-time {\it manifold substantivalism} (Earman \&
Norton, 1987), opened a rich philosophical debate that is still
alive today. The Hole Argument was immediately regarded by virtually
all participants in the debate (Bartels, 1984; Butterfield, 1984,
1987, 1988, 1989; Maudlin, 1988; Rynasiewicz, 1994, 1996, and many
others) as being intimately tied to the deep nature of space and
time, at least as they are represented by the mathematical models of
GR. From 1987 onward, the debate centered essentially about the
ontological position to be taken in interpreting the so-called {\it
Leibniz equivalence}, which is the terminology introduced by Earman
and Norton to characterize philosophically the relation between
diffeomorphic models of GR satisfying the assumptions of the Hole
Argument. It must be acknowledged that until now the debate had a
purely philosophical relevance. Indeed, from the physicists' point
of view, GR has indeed been immunized against the Hole Argument -
leaving aside any underlying philosophical issue - by simply
embodying the Argument in the statement that {\it mathematically
different} solutions of Einstein's equations related by {\it
passive} - as well as {\it active} (see later) - diffeomorphisms are
{\it physically equivalent}. From the technical point of view, the
natural reading of the consequences of the Hole Argument was then
that the mathematical representation of space-time in GR unavoidably
contains {\it superfluous structure}.
\medskip

The main scope of this paper is to show that the immunization
statement quoted above cannot be regarded as the last word on this
matter from both the physical and the philosophical point of view,
and that GR contains in itself the remedy for isolating what seems
to be superfluous structure of the mathematical representation. In
this connection, we wish to remember that, already in 1984, Michael
Friedman was very explicit about the unsatisfactory epistemological
{\it status} of the widespread understanding of the relation between
diffeomorphic models in terms of Leibniz equivalence, when he wrote
({\it our emphasis}):

\begin{quotation}
\footnotesize \noindent Further, if the above models are indeed
equivalent representations of the same situation (as it would seem
they must be) then {\it how do we describe this physical situation
{\it intrinsically}?} Finding such an {\it intrinsic}
characterization (avoiding quantification over {\it bare} points)
appears to be a non-trivial, {\it and so far unsolved mathematical
problem}. (Note that it will not do simply to replace points with
equivalence classes of points: for, in many cases, the equivalence
class in question will contain {\it all} points of the manifold
(Friedman, 1984, p.663.)
\end{quotation}

Friedman's thought was then that the Hole Argument leaves an
unsolved problem about the characterization  of {\it intrinsic
space-time structure}, rather than an ontological question about the
existence of space-time.
Now, we claim that we have solved this problem, in this same spirit,
to the extent in which a degree of intrinsic-ness can be reached in
GR. Clearly, given the enormous mathematical variety of possible
solutions of Einstein's equations one should not expect that a
clarification of Friedman's question is obtained {\it in general}.
We shall indeed conclude that some of the main questions we discuss
can be clarified for the general class of globally hyperbolic
space-times, while some others for a definite continuous family of
generic solutions corresponding to spatially non-compact,
asymptotically-flat space-times (hereafter called by the acronym
C-K)\footnote{The Christodoulou-Klainermann space-times
(Christodoulou \& Klainermann, 1993), which are also privileged from
the point of view of the inclusion of elementary particles.}, but
{\it not} for the spatially compact ones.

\medskip

Conceptually, our solution is developed in three parts:

i) unveiling the physical meaning of Leibniz equivalence and
thereby, through the disclosure of the alleged superfluous
structure, dis-solving the philosophical bearing of the Hole
Argument. This can be done for all the globally hyperbolic models of
GR;

ii) constructing a {\it physical individuation} of point-events in
terms of the {\it autonomous degrees of freedom} of the
gravitational field ({\it Dirac observables}, hereafter called DO).
This can be achieved for the C-K models of GR, and discloses a {\it
highly non-local and holistic} space-time structure. From the
philosophical side, the non-uniformity and dynamical richness of
space-time structure so unveiled lends itself to support a {\it new
structuralist view}\footnote{As already advocated by Mauro Dorato
some years ago (Dorato, 2000).} that we call {\it point
structuralism}. This view appears to be a {\it tertium quid} that
tries in some sense to overcome the crudeness of the historical
debate on the nature of space-time by including elements common to
the traditions of both {\it substantivalism} and {\it
relationism}\footnote{The {\it substantivalist} position is a form
of realism about certain spatiotemporal structures, being committed
to believing in the existence of those entities that are quantified
over by our space-time theories, in particular space-time points. It
conceives space-time, more or less, as a substance, that is as
something that exists independently of any of the things in it. In
particular, the so-called {\it manifold substantivalism} identifies
space-time with the bare mathematical manifold of events glued
together by the topological and differential structure. On the other
hand, the strong {\it relationist} position is the view that
space-time arises as a mere abstraction from the spatiotemporal
properties of other things, so that spatio-temporal relations are
derivative and supervenient on physical relations obtaining among
events and physical objects. Note that both a simple
anti-substantivalist position and a tipical relationist position do
not deny the reality of space-time (they are not merely
anti-realist), but assert that space-time has no reality
independently of the bodies or fields it contains. The crucial
question for our notion of spatiotemporal structuralism, is
therefore the specification of the nature of fields that are {\it
indispensable} for the very definition of physical space-time (e.g.
the gravitational field with its causal structure) as distinguished
from {\it other} physical fields.}. Note, furthermore, that though
conceptually independent of our specific methodology (see later on),
the disclosure of this space-time structure follows from the choice
of the individuation procedure and {\it not} from the Hole Argument
in itself. Concerning the Hole Argument, our analysis of it entails
only a negative philosophical import. Indeed, any attempt to uphold
{\it manifold substantivalism} or {\it any other metaphysical
doctrine} about space-time points, {\it in the face of the Hole
Argument}, becomes definitely irrelevant. Finally:

iii) as a by-product of our analysis concerning the C-K class, we
show a coherent interpretation of the ADM Hamiltonian formulation of
metric gravity in which the so-called {\it weak ADM energy} does
generate {\it real temporal modifications} of the DO. This gives new
evidence to the fact that statements like the assertion of the
so-called {\it frozen-time picture of evolution}, as other
ontological claims about GR, are {\it model-dependent}.
\medskip

The concrete realization of point i) constitutes the conceptual
basis for the development of the whole program. We will show that
physical equivalence of solutions means much more than mere
difference in mathematical description: actually it entails
equivalence of the descriptions of gravitational phenomena given in
different {\it global, non-inertial frames}, which are {\it extended
space-time laboratories} (hereafter called NIF), with their
(dynamically determined) {\it chrono-geometrical conventions} and
{\it inertial potentials}\footnote{A precise definition of NIF is
given in Section IV.}. In fact, in developing our program of
resolution of Friedman's question, we have been naturally led to
rephrase GR in terms of {\it generalized inertial effects}, viz.
those effects which are unavoidably met with by any empirical access
to the theory due to a global consequence of the equivalence
principle. Incidentally, although abandoned later on, the
methodological pre-eminence of non-inertial frames in dealing with
GR was evident in Einstein's original attitude towards gravitation.
Since then, this attitude has never been recovered in the
literature. Note, on the other hand, that today extended
laboratories like GPS cannot avoid the issue of inertial effects and
that contemporary gravitational experiments in space will tend to
get a final clarification of this topic.

\medskip

Technically, this work is based on a full implementation of Dirac's
theory of constraint as applied to metric gravity. The recourse to
the Hamiltonian formalism is necessary for our goals for many
important reasons.

i) The Hamiltonian approach guarantees that the initial value
problem of Einstein's equations is mathematically well-posed, a
circumstance that does not occur in a natural way within the
configurational Lagrangian framework (or "manifold way") because of
the non-hyperbolic nature of Einstein's equations (Friedrich \&
Rendall, 2000; Rendall, 1998). This is a crucial point which we will
come back to in Section VI, with greater technical detail. The Hole
Argument, in fact, is inextricably entangled with the {\it
initial-value problem} of GR, even if it has never been explicitly
discussed in that context in a systematic way\footnote{Actually,
David Hilbert was the first person to discuss the Cauchy problem for
Einstein's equations and to realize its connection to the Hole
phenomenology (see Hilbert, 1917). He discussed the issue in the
context of a general-relativistic generalization of Mie's
special-relativistic nonlinear electrodynamics, and pointed out {\it
the necessity of fixing a special geometrically adapted} ("Gaussian"
in his terms, or geodesic normal as known today) {\it coordinate
system}, to assure the causality of the theory (in this connection
see Howard \& Norton, 1993).}.

ii) In the context of the Hamiltonian formalism, we can resort to
Bergmann and Komar's theory of general coordinate-group symmetries
(see Bergmann \& Komar, 1972) to clarify the significance of active
diffeomorphisms as \emph{dynamical symmetries} of Einstein's
equations. This point also is crucial: to fully understand the role
played by \emph{active} diffeomorphisms in the original
configurational formulation of the Hole Argument, it is necessary to
interpret them as the \emph{manifold-way} counterparts of {\it on
shell}\footnote{We distinguish {\it off-shell} considerations, made
within the Hamiltonian variational framework before restricting to
the dynamical solutions, from {\it on-shell} considerations, made
after such a restriction.} Hamiltonian gauge transformations, which
are \emph{passive} by definition. It will be seen, again, that a
basic misunderstanding on the philosophical bearing of the Hole
Argument follows directly from a loose and non-algorithmic account
of the Cauchy surface as a purely geometrical entity within the
manifold $M^4$ ${}$ ${}$\footnote{Our stance about the content and
the implications of the original Hole Argument contrasts with the
manifestly covariant and generalized attitude towards the Hole
phenomenology expounded by John Stachel in many papers (see e.g.
Stachel \& Iftime, 2005, and references therein). We will come back
to this point and defend our approach in Sections III and VI.}. On
the other hand, we stress that reaching our conclusions within the
Lagrangian formulation would be technically quite awkward if not
impossible, since the Legendre pull-back of the non-point canonical
transformations of the Hamiltonian formulation would require tools
like the infinite-jet bundle formalism.

iii) The most important reason in favor of the Hamiltonian approach
is that, on the basis of the so-called Shanmugadhasan canonical
transformation (Shanmugadhasan, 1973; Lusanna, 1993), this approach
provides a neat distinction between {\it physical observables} (the
four DO) connected to the (two) autonomous degrees of freedom of the
gravitational field, on one hand, and {\it gauge variables}, on the
other. The latter - which express the typical arbitrariness of the
theory and must be fixed ({\it gauge-fixing}) before solving
Einstein's equations for the autonomous degrees of freedom - turn
out to play a fundamental role, no less than the DO, in both
clarifying the real import of the Hole Argument and, even more,
showing the need for skipping the manifestly covariant perspective
in order to get a full understanding of the physical basis of GR. As
said above, the experimental set up for any kind of measurements
made within the GR theoretical framework requires the constitution
of a NIF. This is exactly what is done - chrono-geometrically - by a
{\it complete gauge-fixing} which, in turn, amounts to a {\it
complete breaking of general covariance}. It should be stressed once
more that breaking general covariance is a theoretical necessity for
the procedure of solving Einstein's equation and not a question of
free choice, let alone a drawback. The basic role of the gauge
variables is, therefore, that of specifying the way in which the
generalized inertial effects, typical of any NIF, affect the
intrisic gravitational degrees of freedom described by the four DO.

After the Shanmugadasan transformation, this mechanism realizes a
characteristic functional split of the metric tensor into an {\it
ontic} and an {\it epistemic} part that can be described as follows:
i) the {\it ontic} part, which is constituted by the four DO and
will be found to specify the intrinsic structure of space-time,
describes, physically, the {\it tidal-like} effects\footnote{Note
that, unlike the Newtonian case where tidal forces are entirely
determined by the variation of the action-at-a-distance generated by
the Newton potential of massive bodies on test particles, in GR we
can have tidal forces even in absence of matter, since they are due
to the autonomous degrees of freedom of the gravitational field.};
ii) the {\it epistemic} part, which is encoded in the metric at the
beginning as arbitrary information and is furnished by the gauge
variables, describes, physically, the {\it generalized inertial}
effects and, after a complete gauge-fixing, specifies the way in
which the {\it ontic} component of the metric field manifests itself
in a definite NIF. Genuine gravitational effects are always dressed
in {\it inertial-like appearances} which undergo inertial changes
when going from a given NIF to another NIF.\footnote{We are
perfectly aware that we are here overstating the philosophical
import of terms like {\it ontic} and {\it epistemic} and their
relationships. Nothing, however, hinges on these nuances in what
follows.}.

iv) Finally, an additional important feature of the solutions of
GR dealt with in our Hamiltonian formulation is the following. The
ADM formalism (Arnowitt, Deser \& Misner, 1962) for {\it spatially
compact} space-times without boundary implies that the Dirac
Hamiltonian generates purely harmless gauge transformations, so
that, {\it being zero on the reduced phase space} (see Section
IV), it {\it cannot engender any real temporal change}. This is
the origin of the so-called {\it frozen evolution} description; in
this connection see Earman, 2002, Belot \& Earman, 1999, 2001.
However, in the case of the C-K family of {\it spatially
non-compact} space-times, {\it internal mathematical consistency}
(requiring the addition of the DeWitt surface term to the
Hamiltonian (DeWitt, 1967), see later) entails that the generator
of temporal evolution, namely the (now non-weakly vanishing) Dirac
Hamiltonian, be instead the so-called {\it weak ADM energy}. This
quantity {\it does generate real temporal modifications} of the
DO.

\medskip

The {\it dual role} of the metric field discussed above highlights
the fact that, while from the mathematical point of view of the
constrained Hamiltonian formalism, GR is a theory like any other
(e.g., electromagnetism and Yang-Mills theory), from the {\it
physical} point of view it is {\it radically different}.
Technically, this can be traced to general covariance itself, i.e.
to the invariance under a group of diffeomorphisms acting on
space-time, instead of invariance under the action of a local {\it
inner} Lie group, like in standard gauge theories. However,
physically, matters are much more complex. First of all, in GR we
cannot rely from the beginning on empirically validated,
gauge-invariant dynamical equations for the {\it local} fields, as
it happens with electro-magnetism, where Maxwell equations can be
written in terms of the gauge-invariant electric and magnetic
fields. On the contrary, Einstein's general covariance (viz. the
basis of the gauge freedom of GR) is such that the introduction of
extra ({\it gauge}) variables does make the {\it mathematical}
description of the geometrical aspects of GR mathematically handy
and elegant, but, by ruling out any background structure at the
outset, it also makes its physical interpretation more intriguing.
In GR the distinction between what is observable and what is not, is
unavoidably entangled with the constitution of the very {\it stage},
space--time, where the play of physics is enacted: a stage, however,
which also takes an active part in the play. In other words, the
gauge-fixing mechanism also plays the {\it double role} of making
the dynamics {\it unique} (as in all gauge theories), and of fixing
the {\it appearance} of the spatiotemporal dynamic background.
Summarizing, as it will be explained, for both the vacuum case and
the case with matter fields, the gauge-fixing (with the correlated
break of general covariance) completes the structural properties of
the general-relativistic space-time. Such fixing is necessary to
solve Einstein's equations, reconstruct the four-dimensional dynamic
chrono-geometry emerging from the initial values of the {\it four
Dirac observables}, and allow empirical access to the theory through
the definition of a dynamically-selected NIF.

At the end of our technical developments, it should be clear why
people (philosophers, especially) should free themselves from the
bewitching fascination of manifest general covariance in order to
fully understand the subtle hindrances underlying Friedman's
question. For general covariance represents a horizon of \emph{a
priori} possibilities for the physical constitution of the
space-time, possibilities that must be actualized in non-covariant
form within any given solution of the dynamical equations.
\medskip

The overall perspective emerging from our analysis amounts to a new
way of looking at the various aspects of the issue of the {\it
objectivity} of the space-time properties of GR. This will be
discussed in detail in Section VI. It will be argued there that the
issue of objectivity must be called into question not only for the
particular case of point-events, but also for many of the basic
spatiotemporal features of the theory, such as causal structure,
one-way velocity of light (see Alba \& Lusanna, 2005a) and the like.
Concerning point events, we shall defend the thesis that, as {\it
relata} within a {\it structure} they do exist and possess some
abstract kind of {\it intrinsic} properties. However, their {\it
physical} properties are {\it relational} being {\it conferred} on
them in a {\it holistic and non-local} way by the whole structure of
the metric field and the extrinsic curvature on a simultaneity
hyper-surface. In a definite physical sense, point-events are
literally {\it identifiable} with the local values of the autonomous
degrees of freedom of the gravitational field (DO). In this way both
the metric field and the point-events maintain - to paraphrase
Newton - their {\it own manner of existence} and this justifies our
terminology of {\it "point-structuralism"}. Furthermore, this
conception does not dissolve physical entities into mathematical
structures, so that it supports a moderate {\it entity-realist}
attitude towards both the metric field and its point-events, as well
as a {\it theory-realist} attitude towards Einstein's field
equations.  However, the degree of {\it objectivity} that, on the
basis of our solution of Friedman's question, should be attributed
to the {\it physical individuation} of point-events as well as to
other basic structures of space-time is a matter of discussion. This
work should be considered a case study for the defence of a thesis
about the physical nature of point-events and other important
spatiotemporal elements in certain classes of models of GR. Our
conclusion will be that all these structures maintain - in a
definite sense - a {\it weak form of physical objectivity}.

\bigskip

Although great part of the technical developments underlying this
work have already been treated elsewhere (Pauri \& Vallisneri, 2002;
Lusanna \& Pauri, 2006-I, 2006-II, hereafter denoted by LPI and
LPII, respectively, where additional properties of the
Christodoulou-Klainermann family of space-times are also discussed),
some technical elements are introduced here for the first time. For
a more general philosophical presentation, see Dorato \& Pauri,
2004.

Section II is devoted to a synopsis of Noether and dynamical
symmetries of GR within the configurational description in a
mathematical manifold $M^4$, together with a brief presentation of
the $Q$ group of Bergmann \& Komar (1972). This is the largest group
of {\it passive dynamical symmetries} of Einstein's equations and is
instrumental to our understanding of the Hole Argument. The latter
is expounded in detail in Section III. The basic ingredients of the
ADM formulation as applied to the Christodoulou-Klainermann family
of space-times and its canonical reduction, the chrono-geometric
meaning of the gauge-fixings, and the constitution of the NIF are
discussed in Section IV, together with the issue of temporal change.
In Section V we show how the {\it ontic} part of the metric (the
autonomous degrees of freedom of the gravitational field) may confer
a {\it physical individuation} of space-time points\footnote{There
is an unfortunate ambiguity in the usage of the term {\it space-time
points} in the literature: sometimes it refers to elements of the
mathematical structure that is the first layer of the space-time
model, other times to the points interpreted as {\it physical}
events. We will adopt the term {\it point--event} in the latter
sense and simply {\it point} in the former.}. In Section VI we take
up the results obtained in the previous Sections and re-discuss them
with respect to the issue of the objectivity of space-time
structures in general. The concluding remarks are devoted to a
philosophical assessment of our results in view of the traditional
debate between {\it substantivalism} and {\it relationism}, as well
as in view of some more recent discussions about {\it structural
realism}.

\section{Noether and dynamical symmetries}

Standard general covariance, which essentially amounts to the
statement that Einstein's equations for the metric field
${}^4g(x)$ have a tensor character, implies first of all that the
basic equations are form invariant under general coordinate
transformations ({\it passive} diffeomorphisms), so that the
Lagrangian density in the Einstein-Hilbert action is singular.
Namely, passive diffeomorphisms {\it are local Noether symmetries}
of the action, so that Dirac constraints appear correspondingly in
the Hamiltonian formulation. The singular nature of the
variational principle of the action entails in turn that four of
Einstein's equations be in fact {\it Lagrangian constraints},
namely restrictions on the Cauchy data, while four combinations of
Einstein's equations and their gradients vanish identically ({\it
contracted Bianchi identities}). Thus, the ten components of the
solution ${}^4g_{\mu\nu}(x)$ are in fact functionals of only two
"deterministic" dynamical degrees of freedom and eight further
degrees of freedom which are left {\it completely undetermined} by
Einstein's equations {\it even once the Lagrangian constraints are
satisfied}. This state of affairs makes the treatment of both the
Cauchy problem of the non-hyperbolic system of Einstein's
equations and the definition of observables within the Lagrangian
context (Friedrich \& Rendall, 2000; Rendall, 1998) extremely
complicated.

For the above reasons, standard general covariance is then
interpreted, in modern terminology, as the statement that {\it a
physical solution of Einstein's equations} properly corresponds to
a {\it 4-geometry}, namely the equivalence class of all the
4-metric tensors, solutions of the equations, written in all
possible 4-coordinate systems. This equivalence class is usually
represented by the quotient ${}^4Geom = {}^4Riem / {}_PDiff\,
M^4$, where ${}^4Riem$ denotes the space of metric tensor
solutions of Einstein's equations and ${}_PDiff\ $ is the infinite
group of {\it passive} diffeomorphisms (general coordinate
transformations). On the other hand, any two {\it inequivalent}
Einstein space-times are different 4-geometries or "universes".

Consider now the abstract differential-geometric concept of {\it
active} diffeomorphism $D_A$ and its consequent action on the tensor
fields defined on the differentiable manifold $M^4$ [see, for
example, (Wald, 1984, pp.438-439)]. An {\it active} diffeomorphism
$D_A$ maps points of $M^4$ to points of $M^4$: $D_A: p {\rightarrow
\hspace{.2cm}}\ p' = D_A \cdot p$. Its tangent map $D_A^{*}$ maps
tensor fields $T {\rightarrow \hspace{.2cm}} D_A{*} \cdot T$ in such
a way that $[T](p) {\rightarrow \hspace{.2cm}} [D_A^{*} \cdot T](p)
\equiv [T^{'}](p)$. Then $[D_A^{*} \cdot T](p) = [T](D_A^{-1}\cdot
p)$. It is seen that the transformed tensor field $D_A^{*} \cdot T$
is a {\it new} tensor field whose components in general will have at
$p$ values that are {\it different} from those of the components of
$T$. On the other hand, the components of $D_A^* \cdot T$ have at
$p'$ - by construction - the same values that the components of the
original tensor field $T$ have at $p$: $T^{'}(D_A \cdot p) = T(p)$
or $T'(p) = T(D_A^{-1}\cdot p)$. The new tensor field $D_A^* \cdot
T$ is called the {\it drag-along} (or {\it push-forward}) of $T$.
There is another, non-geometrical - so-called {\it dual} - way of
looking at the active diffeomorphisms (Norton, 1987). This {\it
duality} is based on the circumstance that in each region of $M^4$
covered by two or more charts there is a one-to-one correspondence
between an {\it active} diffeomorphism and a specific coordinate
transformation. The coordinate transformation ${\cal T}_{D_A}: x(p)
{\rightarrow \hspace{.2cm}}\ x'(p) = [{\cal T}_{D_A}x](p)$ which is
{\it dual} to the active diffeomorphism $D_A$ is defined so that
$[{\cal T}_{D_A}x](D_A \cdot p) = x(p)$. Essentially, this {\it
duality} transfers the functional dependence of the new tensor field
in the new coordinate system to the old system of coordinates. By
analogy, the coordinates of the new system $[x']$ are said to have
been {\it dragged-along} with the {\it active} diffeomorphism $D_A$.
It is important to note here, however, that the above {\it dual
view} of active diffeomorphisms, as particular {\it
coordinate}-transformations, is only {\it implicitly} defined for
the moment .

In the abstract coordinate-independent language of differential
geometry, Einstein's equations for the vacuum

\bea
 {}^4G_{\mu\nu}(x)\,\, {\buildrel {def}\over =}\,\, {}^4R_{\mu\nu}(x)
- {1\over 2}\, {}^4R(x)\, {}^4g_{\mu\nu}(x)  = 0.
 \label{E1}
  \eea

\noindent can be written as $G = 0$, where $G$  is the Einstein
2-tensor ($G = G_{\mu\nu}(x)\, dx^{\mu} \bigotimes dx^{\nu}$ in the
coordinate chart $x^{\mu}$). Under an {\it active diffeomorphism}
$D_A: M^4 \mapsto M^4$, $D_A \in {}_ADiff\, M^4$, we have $G = 0
\mapsto D^*_A\, G = 0$, which shows that active diffeomorphisms are
{\it dynamical symmetries} of Einstein's tensor equations, i.e. {\it
they map solutions into solutions}.

We have clarified elsewhere (LPI) the {\it explicit}
relationships\footnote{At least for the infinitesimal active
transformations.} existing between passive and active
diffeomorphisms on the basis of an important paper by Bergmann and
Komar (1972), in which it is shown that {\it the largest group of
passive dynamical symmetries of Einstein's equations} is not
${}_PDiff\, M^4$ [$x^{{'}\, \mu} = f^{\mu}(x^{\nu})$], but rather a
larger group of transformations of the form

\bea
 Q:&& x^{{'}\, \mu} = f^{\mu}(x^{\nu}, {}^4g_{\alpha\beta}(x)),
 \nonumber \\
 &&{}\nonumber \\
  {}^4g^{'}_{\mu\nu}(x^{'}(x)) &=& {{\partial h^{\alpha}(x^{'},
  {}^4g^{'}(x^{'}))}\over {\partial
 x^{'\, \mu}}}\, {{\partial h^{\beta}(x^{'}, {}^4g^{'}(x^{'}))}\over
 {\partial x^{'\, \nu}}}\, {}^4g_{\alpha\beta}(x).
 \label{E2}
 \eea
\medskip

In the case of completely Liouville-integrable systems, dynamical
symmetries can be re-interpreted as maps of the space of Cauchy
data onto itself. Although we do not have a general proof of the
integrability of Einstein's equations, we know that if the initial
value problem is well-posed and defined\footnote{It is important
to stress that in looking for global solutions of Einstein's
equations as a system of partial differential equations, a number
of preliminary specifications must be given. Among other things:
a) the topology of space-time; b) whether the space-time is
spatially compact or asymptotically flat at spatial infinity; c)
whether or not in the spatially compact case there is a spatial
boundary; d) the nature of the function space and the class of
boundary conditions for the 4-metric and its derivatives, either
at spatial infinity or on the spatial boundary (only in the
spatially compact case without boundary there is no need of
boundary conditions, replaced by periodicity conditions, so that
these models of GR show the well-known Machian aspects which
influenced Einstein and Wheeler). After these specifications have
been made, an {\it off-shell model} of GR is identified. What
remains to be worked out is the characterization of a well-posed
initial value problem. Modulo technicalities, this requires
choosing a 4-coordinate system and finding which combinations of
the equations are of {\it elliptic} type (restrictions on the
Cauchy data) and which are of {\it hyperbolic} type (evolution
equations), namely the only ones requiring an initial value
problem. At the Hamiltonian level, the elliptic equations are the
first-class constraints identifying the constraint sub-manifold of
phase space (see Section IV), while the hyperbolic equations are
the Hamilton equations in a fixed gauge (a completely fixed
Hamiltonian gauge corresponds {\it on-shell} to a 4-coordinate
system, see Section IV). When the gauge variables can be separated
from the Dirac observables, only the latter need an initial value
problem (the gauge variables are {\it arbitrary}, modulo
restrictions upon their range coming from the structure of the
gauge orbits inside the constraint sub-manifold). Finally, given a
space-like Cauchy surface in a 4-coordinate system (or in a fixed
Hamiltonian gauge), {\it each admissible set of Cauchy data gives
rise to a different "universe" with the given boundary
conditions}. Clearly, each universe is defined modulo passive
diffeomorphisms changing both the 4-coordinate system and the
Cauchy surface (or modulo the Hamiltonian gauge transformations
changing the gauge and the Cauchy surface) and also modulo the
({\it on-shell}) active diffeomorphisms.}, as it is in the ADM
Hamiltonian description, {\it the space of Cauchy data is
partitioned in gauge-equivalent classes of data}: all of the
Cauchy data in a given class identify a single 4-geometry or {\it
"universe"}. Therefore, under the given hypothesis, {\it the
dynamical symmetries of Einstein's equations fall into two classes
only}: a) those mapping different {\it "universes"} among
themselves, and b) those acting within a single Einstein {\it
"universe"}, mapping gauge-equivalent Cauchy data among
themselves. It is remarkable that, at least for the subset $Q'
\subset Q$ (passive counterpart of a subset ${}_ADiff^{'}\, M^4
\subset {}_ADiff\, M^4$) which corresponds to {\it mappings among
gauge-equivalent Cauchy data}, the transformed metrics do indeed
belong to the {\it same 4-geometry}, i.e. the same equivalence
class generated by applying all {\it passive diffeomorphisms} to
the original 4-metrics: $ {}^4Geom  = {}^4Riem / {}_PDiff\, M^4 =
{}^4Riem / Q'$\footnote{Note, incidentally, that this circumstance
is mathematically possible only because ${}_PDiff\, M^4$ is a {\it
non-normal dense} sub-group of $Q^{'}$.}.

\medskip

Note finally that: a) an {\it explicit} passive representation of
the infinite group of ${}_ADiff\, M^4$ is necessary anyway for our
Hamiltonian treatment of the Hole Argument, as well as for any
comparison of the various viewpoints existing in the literature
concerning the {\it solutions} of Einstein's equations; b) the
group $Q$ describes the dynamical symmetries of Einstein's
equations which are {\it also} local Noether symmetries of the
Einstein-Hilbert action.

\medskip

In conclusion, what is known as a {\it 4-geometry}, is also an
equivalence class of solutions of Einstein's equations {\it modulo}
the subset of dynamical symmetry transformations ${}_ADiff'\, M^4$,
whose passive counterpart is $Q^{'}$. Therefore, following Bergmann
\& Komar (1972), Wald (1984), we can state \footnote{Eqs.(2.3) are
usually taken for granted in mathematical physics, at least at the
heuristic level. Since, however, the control in large of the group
manifold of infinite-dimensional groups like ${}_PDiff\, M^4$ and
$Q^{'}$ is, as yet, an open mathematical issue, one cannot be more
rigorous on this point: see also the end of Section III. For more
details about these issues, the interested reader should see LPI and
LPII, where a new subset ${}Q_{can}$ of $Q$ is introduced, namely
the Legendre pullback of the {\it on-shell} Hamiltonian canonical
transformations. In LPI and LPII it is shown that it also holds $
{}^4Geom = {}^4Riem / Q_{can}$, since, modulo technicalities, we
have ${}Q_{can} = Q'$. Note that ${}_PDiff\, M^4 \cap Q_{can}$ are
the {\it passive diffeomorphisms which are re-interpretable as
Hamiltonian gauge transformations}.}

\bea
 {}^4Geom = {}^4Riem / {}_PDiff\, M^4 = {}^4Riem / Q' = {}^4Riem /
 {}_ADiff'\, M^4.
 \label{E3}
 \eea

\section{The Hole Argument and its dis-solution: a first look}

Although the issue could not be completely clear to Einstein in
1916, as shown by Norton  (1987, 1992, 1993), it is precisely the
nature of dynamical symmetry of the {\it active diffeomorphisms}
that has been considered as expressing the {\it physically
relevant} content of {\it general covariance}, as we shall
presently see.

Remember, first of all, that a {\it mathematical model} of GR is
specified by a four-di\-men\-sio\-nal mathematical manifold $M^4$
and by a metrical tensor field $g$, where the latter  represents
{\it both} the chrono-geometrical structure of space-time {\it
and} the potential for the inertial-gravitational field.
Non-gravitational physical fields, when they are present, are also
described by dynamical tensor fields, which appear to be sources
of Einstein's equations. Assume now that $M^4$ contains a {\it
hole} $\mathcal{H}$: that is, an open region where all the
non-gravitational fields vanish so that {\it the metric obeys the
homogeneous Einstein equations}. On $M^4$ we can define an {\it
active} diffeomorphism $D_{A}$ that re-maps the points inside
$\mathcal{H}$, but blends smoothly into the identity map outside
$\mathcal{H}$ and on the boundary. By construction, for any point
$x \in \mathcal{H}$ we have (in the abstract tensor notation)
$g'(D_A x)= g(x)$, but of course $g'(x) \neq g(x)$ (in the same
notation). The crucial fact is that from the general covariance of
Einstein's equations it follows that if $g$ is one of their
solutions, so is the {\it drag-along} field $g' \equiv D^{*}_{A}
g$.

What is the correct interpretation of the new field $g'$? Clearly,
the transformation involves an {\it active redistribution of the
metric over the points of the manifold in $\mathcal{H}$}, so the
critical question is whether and how the points of the manifold are
primarily {\it individuated}. Now, {\it if} we think of the points
of $\mathcal{H}$ as {\it intrinsically} {\it individuated} {\it
physical events}, where {\it intrinsic} means that their identity is
autonomous and independent of any physical field, the metric in
particular - a claim that is associated with any kind of {\it
manifold substantivalism} - then $g$ and $g'$ must be regarded as
{\it physically distinct} solutions of Einstein's equations (after
all, $g'(x) \neq g(x)$ at the {\it same} point $x$). This appeared
as a devastating conclusion for the causality of the theory, because
it implied that, even after we specify a physical solution for the
gravitational and non-gravitational fields outside the hole - in
particular, {\it on a Cauchy surface for the initial value problem}
- we are still unable to predict a unique physical solution within
the hole.

According to Earman and Norton (1987), the way out of the Hole
Argument lies in abandoning {\it manifold substantivalism}: they
claim that if diffeomorphical\-ly related metric fields were to
represent different physically possible "universes", then GR would
turn into an {\it indeterministic} theory. And since the issue of
whether determinism holds or not at the {\it physical} level cannot
be decided by opting for a {\it metaphysical} doctrine like {\it
manifold substantivalism}, they conclude that one should refute any
kind of such {\it substantivalism}. Since, however, relationism does
not amount to the mere negation of substantivalism, and since the
literature contains so many conflicting meanings of the term
"relationism", they do not simply conclude that space-time is
relational. They state the more general assumption (which - they
claim - is applicable to all space-time theories) that
"diffeomorphic models in a space-time theory represent the same
physical situation", i.e. must be interpreted as describing the same
{\it "universe"} ({\it Leibniz equivalence}).

The fact that the Leibniz equivalence seems here no more than a
sophisticated re-phrasing of what physicists consider a foregone
conclusion for general relativity, should not be taken at face
value, for the real question for an opposing "weak substantivalist"
is whether or not space-time should be simply identified with the
bare manifold deprived of any physical field, and of {\it the metric
field} in particular, as Earman and Norton do, instead of with a set
of points each endowed with its own {\it metrical
fingerprint}\footnote{See, for example, Bartels, 1994 and Maudlin,
1988.}. Actually, this "weak substantivalist" could sustain the
conviction - as we ourselves do - that the metric field, because of
its basic {\it causal structure}, has ontological priority (Pauri,
1996) over all other fields and, therefore, it is {\it not} like any
other field, as Earman and Norton would have it.

We do believe that the bare manifold of points, deprived of the
infinitesimal Pythagorean structure defining the basic distinction
between temporal and spatial directions, let alone the causal
structure which teaches all the other fields how to move, can hardly
be seen as space-time. We insist, therefore, that in order to be
able to speak of space-time the definition of a metric field is a
necessary condition. Consequently, and in agreement with Stachel
(1993), we believe that asserting that $g$ and $D^{*}_{A} g$
represent {\it one and the same gravitational field} implies that
{\it the mathematical individuation} of the points of the
differentiable manifold by their coordinates {\it has no physical
content unless a metric tensor is specified} \footnote{{\it
Coordinatization} is the only way to individuate the points {\it
mathematically} since, as stressed by Hermann Weyl: ''There is no
distinguishing objective property by which one could tell apart one
point from all others in a homogeneous space: at this level,
fixation of a point is possible only by a {\it demonstrative act} as
indicated by terms like {\it this} and {\it there}.'' (Weyl, 1946,
p. 13). See also Schlick, 1917, quoted in M.Friedman, 2003, p.165.}.
Stachel stresses that if $g$ and $D^{*}_{A}g$ must represent the
same gravitational field, they cannot be physically distinguished in
any way. Accordingly, when we act on $g$ with $D^{*}_{A}$ to create
the {\it drag-along} field $D^{*}_{A} g $, no element of physical
significance can be left behind: in particular, {\it nothing} that
could identify a point $x$ of the manifold itself as the {\it same
point} of space-time for both $g$ and $D^{*}_{A} g$. Instead, when
$x$ is mapped onto $x' = D^{*}_{A}x$, it {\it carries over its
identity}, as specified by $g'(x')= g(x)$. This means,  for one
thing, that "the last remnant of physical objectivity" of space-time
points, if any, should be sought for in the physical content of the
metric field itself. Note, however, that while this point of view
about the Hole Argument casts new light upon the fact that $g$ and
$g'$ must represent the same physical situation, it still does not
explain the source of the superfluous structure of the mathematical
representation and, therefore, not even the origin of the prima
facie {\it difference} between $g$ and $g'$.

Anyway, these remarks led Stachel to the important conclusion that
{\it vis \'{a} vis} the physical point-events, the metric actually
plays the role of {\it individuating field}. Precisely, Stachel
suggested that this individuating role could be implemented by four
invariant functionals of the metric, which Komar and Bergmann (Komar
1958; Bergmann \& Komar, 1960) had already considered. Stachel,
however, did not follow up this proposal by providing a concrete
realization in terms of solutions of Einstein's equations, something
that we instead will presently do. At the same time we will show in
Section VI that Stachel's suggestion, as it stands, remains at a too
abstract level and fails to exploit the crucial distinction between
{\it ontic} and arbitrary {\it epistemic} content of the
Bergmann-Komar invariant functionals of the metric. This content, in
fact, must be specified in order to calculate the functionals on the
solutions.

\medskip

We conclude this Section by summarizing the implications of our
analysis of the meaning and philosophical import of the Hole
Argument. The force of the {\it indeterminacy argument} apparently
rests on the following basic facts: (i) a solution of Einstein's
equations must be {\it preliminarily} individuated outside (and, of
course, inside) the Hole, otherwise there would be no 'meat' for the
Argument itself. Although the original formulation of the Hole
Argument, as well as many subsequent expositions of it, are silent
on this point, we will see that the Hole Argument is unavoidably
entangled with the initial value problem; (ii) the {\it active}
diffeomorphism $D^{*}_{A}$, which is purportedly chosen to be the
{\it identity} outside the {\it hole} $\mathcal{H}$, is a {\it
dynamical symmetry} of Einstein's equations, so that it maps
solutions onto solutions, which in general will be equivalent (as
4-geometries or Einstein "universes") or not ; (iii) since
$D^{*}_{A}$ is, by hypothesis, the identity on the Cauchy
hyper-surface, it cannot map a solution defining a given Einstein
"universe" onto a different "universe", {\it which would necessarily
correspond to inequivalent Cauchy data}; but (iv) nevertheless, we
are told by the Hole Argument that $D^{*}_{A}$ engenders a
"different" solution inside the Hole.

As we shall later see with greater evidence, in spite of the {\it
prima facie} geometric obviousness of the identity condition
required for $D^{*}_{A}$ outside the Hole, it is quite illusory to
try to explain all the facets of the relations of the Argument
with the initial value problem in the purely abstract way of
differential geometry. The point is that the configurational
four-dimensional formulation cannot exploit the advantage that the
Hamiltonian formulation possesses of working {\it off-shell}. This
is crucial since in the present context the four-dimensional
active diffeomorphisms - {\it qua} dynamical symmetries of
Einstein's equations - must be directly applied to {\it solutions}
of GR. These solutions, however, cannot be exhaustively managed in
the four-dimensional configurational approach in terms of initial
data because of the non-hyperbolicity of Einstein's equations
\footnote{The reader should avoid the impression that our
criticisms about the weakness of the configurational approach be
dictated by personal preferences for Hamiltonian methods or be
simple questions of taste. The issue is in fact very important. In
LP2 we gave a detailed analysis of the drawbacks one can encounter
by adopting trusting confidence in a mixing of Lagrangian and
geometric considerations involving a non-algorithmic attitude
towards the Cauchy surface. A brief reconsideration of this
analysis is given in Section VI.}. The Hole Argument needs the
Cauchy problem to be formulated outside the Hole explicitly and
{\it in advance}, a fact that requires abandoning the Lagrangian
way in favor of Hamiltonian methods. At this point, the results of
the previous Section (the passive counterpart of $D_{A}$ must
belong to $Q'$, or belong to $Q$ but not to $Q'$) leave us with
the sole option that, {\it once considered in view of the passive
Hamiltonian description}, the {\it active} diffeomorphism
$D^{*}_{A}$ exploited by the Argument must lie in the subset $Q'$
(${}_ADiff^{'}\ M^4$). Then, however, {\it it must necessarily map
Cauchy data onto gauge-equivalent Cauchy data}, precisely those
gauge-equivalent data that generate the allegedly "different"
solution within the Hole. In the end, the "difference" will turn
out to correspond to a mere {\it different choice of the gauge}
for the {\it same solution}. Thus {\it Leibniz equivalence} boils
down to mere {\it gauge equivalence} in its strict sense
\footnote{The physical meaning of this equivalence will be
clarified in Section IV.}, an effect that - let us stress it again
- cannot be transparently displayed in the configurational
geometric description. On the other hand, were the {\it active}
diffeomorphism $_{A}$, once passively rephrased, to belong to the
group $Q$ but not to the subset $Q^{'}$ (i.e. were it originally
lying in ${}_ADiff\, M^4 $, but not in ${}_ADiff^{'}\, M^4)$),
then it would not correspond to a mere gauge equivalence and it
would necessarily modify the Cauchy data outside the Hole.
Therefore it would lead to a really different Einstein "universe"
but it would violate the assumption of the Hole Argument that
$D_{A}$ be the identity on the Cauchy hyper-surface. In any case,
it is seen that the disappearance of the "indeterminacy" rests
upon the necessity of formulating the Cauchy problem {\it before}
talking about the relevant properties of the solutions.

We conclude that - to the extent that the Cauchy problem is
well-posed, i.e. {\it in every globally hyperbolic space-time} and
not necessarily in the C-K class only - exploiting the original Hole
Argument to the effect of asking ontological questions about the
general relativistic space-time is an enterprise devoid of real
philosophical impact, in particular concerning the menace of
indeterminism. There is clearly no room left for upholding "manifold
substantivalism", "different worlds", "metric essentialism" or any
other metaphysical doctrine about space-time points {\it in the face
of the Hole Argument}. Of course, such metaphysical doctrines could
still be defended, yet independently of the Hole Story.

\section{Christodoulou-Klainermann
space-times, 3+1 splitting, and ADM canonical reduction}

\noindent The Christodoulou-Klainermann space-times are a
continuous family of space-times that are non-compact, globally
hyperbolic, asymptotically flat at spatial infinity (asymptotic
Minkowski metric, with asymptotic Poincar\'e symmetry group) and
topologically trivial ($M^4 \equiv R^3 \times R$), supporting
global 4-coordinate systems.

The ADM  Hamil\-to\-nian ap\-pro\-ach  starts with a 3+1 splitting
of the 4-dimensional manifold $M^4$ into constant-time
hyper-surfaces $\Sigma_{\tau} diffeomorphic to R^3$, indexed by
the {\it parameter time} $\tau$, each equipped with coordinates
$\sigma^a$ (a = 1,2,3) and a three-metric ${}^3g$ (in components
${}^3g_{ab}$). The {\it parameter time} $\tau$ and the coordinates
$\sigma^a$ (a = 1,2,3) are in fact {\it Lorentz-scalar, radar}
coordinates {\it adapted} to the 3+1 splitting (Alba \& Lusanna,
2003, 2005a). They are defined with respect to an arbitrary, in
general accelerated, observer, a centroid $X^{\mu}(\tau )$, chosen
as origin of the coordinates, whose proper time may be used as the
parameter $\tau$ labelling the hyper-surfaces. On each
hyper-surface {\it all the clocks are conventionally synchronized
to the value $\tau$}. The simultaneity (and Cauchy) hyper-surfaces
$\Sigma_{\tau}$ are described by the embedding functions $x^{\mu}
= z^{\mu}(\tau , \sigma^a) = X^{\mu}(\tau ) + F^{\mu}(\tau ,
\sigma^a)$, $F^{\mu}(\tau , 0^a) = 0$.

All this machinery builds up, at the chrono-geometric level, a {\it
global, extended frame of reference}, realizing a {\it non-rigid,
non-inertial, laboratory} (the only one existing in GR due to the
equivalence principle)\footnote{All the details of this structure
can be found in LPI, where it is shown in particular how two
congruences of time-like observers are naturally associated to a NIF
.}. This global laboratory will be called a NIF. As we shall
presently show, any {\it chrono-geometrically possible} NIF is the
result of a {\it complete gauge-fixing}, a procedure that determines
the {\it appearance of gravitational phenomena} by uniquely
specifying {\it the form} of the inertial forces (Coriolis, Jacobi,
centrifugal,...) in each point of a NIF. A crucial difference of
this structure in GR with respect to the Newtonian case (see also
footnote 10), besides the fact that inertial effects are unavoidable
(they cannot be traced here to "apparent forces" in that they cannot
be removed by a choice of reference frame), is that the inertial
potentials may also depend upon {\it generalized tidal effects} in
addition to the coordinates of the non-inertial frame\footnote{Let
us stress that the {\it scalar radar coordinates} are {\it
intrinsically frame-dependent} since they parametrize a NIF centered
on the arbitrary observer. Furthermore, they are not ordinary
coordinates $x^{\mu}$ in a chart of the Atlas ${\cal A}$ of $M^4$.
They should be properly called {\it pseudo-coordinates} in a chart
of the Atlas ${\cal \tilde A}$ defined by adding to $M^4$ the
extra-structure of all its admissible 3+1 splittings: actually the
new coordinates are {\it adapted} to this extra-structure. If the
embedding of the constant-time hyper-surfaces $\displaystyle
\Sigma_{\tau}$ of a 3+1 splitting into $M^4$ is described by the
functions $\displaystyle z^{\mu}(\tau ,\sigma^a)$, then the
transition functions from the {\it adapted radar-coordinates}
$\sigma^A = (\tau; \sigma^a)$ to the ordinary coordinates are
$\frac{\displaystyle \partial z^\mu(\tau, \sigma^a)}{\displaystyle
\partial \sigma^A}$.}.

An important point to be kept in mind is that the {\it explicit
functional form} of the embedding functions and - consequently - of
the chrono-geometry of the 3 + 1 splitting of $M^4$, thought to be
implicitly given at the outset, remains undefined until the solution
of Einstein's equations is worked out in a fixed gauge. Likewise, it
is only after the solution emerges from given initial data of the
four DO in that gauge that a subset of the {\it chrono-geometrically
possible} NIF becomes {\it dynamically selected} NIF together with
their dynamically determined "conventions": see later on.

\medskip

Now, start at a point on $\Sigma_{\tau}$, and displace it
infinitesimally in a direction that is normal to $\Sigma_{\tau}$.
The resulting change in $\tau$ can be written as $\triangle\,
\tau$ = $N d\tau$, where N is the so-called {\it lapse function}.
Moreover, the time displacement $d\tau$ will also shift the
spatial coordinates: $\sigma^{a}(\tau + d\tau)$ = $\sigma^a(\tau)+
N^{a}d\tau$, where $N^a$ is the {\it shift vector}. Then the
interval between $(\tau,\sigma^{a})$ and $(\tau + d\tau,
\sigma^{a} + d\sigma^{a})$ is: $ds^2 = N^2d\tau^2 -
{}^3g_{ab}(d\sigma^a + N^a d\tau)(d\sigma^b + N^b d\tau)$. The
{\it configurational} variables $N$, $N^{a}$, ${}^3g_{ab}$
(replacing the 4-metric $g$) together with their 10 conjugate
momenta, index a 20-dimensional phase space\footnote{Of course,
all these {\it variables} are in fact {\it fields}.}. Expressed
({\it modulo} surface terms) in terms of the ADM variables, the
ADM action is a function of $N$, $N^{a}$, ${}^3g_{ab}$ and their
first time-derivatives, or equivalently of $N$, $N^{a}$,
${}^3g_{ab}$ and the extrinsic curvature ${}^3K_{ab}$ of the
hyper-surface $\Sigma_{\tau}$, considered as an embedded manifold.

Since Einstein's original equations are not hyperbolic, it turns out
that the canonical momenta are not all functionally independent, but
satisfy four conditions known as {\it primary} constraints (they are
given by the vanishing of the {\it lapse} and {\it shift} canonical
momenta). Another four {\it secondary} constraints arise when we
require that the primary constraints be preserved through evolution
(the secondary constraints are called the {\it super-hamiltonian}
$\mathcal{H}_0 \approx 0 $, and the {\it super-momentum}
$\mathcal{H}_a \approx 0, \ (a = 1,2,3)$ constraints, respectively).
The eight constraints are given as functions of the canonical
variables that vanish on the {\it constraint surface}. The existence
of such constraints implies that not all the points of the
20-dimensional phase space physically represent meaningful states:
rather, we are restricted to the {\it constraint surface} where all
the constraints are satisfied, i.e. to a 12-dimensional (20 - 8)
{\it surface} which, however, does not possess the geometrical
structure of a true phase space. When used as generators of
canonical transformations, the eight constraints map points on the
constraint surface to points on the same surface; these
transformations are known as {\it gauge transformations}.

In order to obtain the correct dynamics for the constrained system,
we must consider the Dirac Hamiltonian, which is the sum of the
DeWitt surface term (DeWitt, 1967) \footnote{The DeWitt surface term
is {\it uniquely} determined as the sum of two parts: a) the surface
integral to be extracted from the Einstein-Hilbert action to get the
ADM action; b) a surface integral due to an integration by parts
required by the Legendre transformation from the ADM action to phase
space [see (Lusanna, 2001) after Eq.(5.5) and (Hawking \& Horowitz,
1996)]. By adding a surface term different from the ADM one, we
would get another action with the same equations of motion but an
a-priori different canonical formulation. Still another option is to
consider the metric and the Christoffel connection as independent
configuration variables: this is the first-order Palatini formalism,
which has much larger gauge freedom, also including second class
constraints. All these canonical formulations must lead anyway to
the same number of physical degrees of freedom.} [present only in
spatially non-compact space-times and becoming the ADM energy after
suitable manipulations (Lusanna, 2001; DePietri \& Lusanna \&
Martucci \& Russo, 2002)], of the {\it secondary} constraints
multiplied by the lapse and shift functions, and of the {\it
primary} constraints multiplied by arbitrary functions (the
so-called {\it Dirac multipliers}). If, following Dirac, we make the
reasonable demand that the evolution of all {\it physical variables}
be unique - otherwise we would have real physical variables that are
indeterminate and therefore neither {\it observable} nor {\it
measurable} - then the points of the constraint surface lying on the
same {\it gauge orbit}, i.e. linked by gauge transformations, must
describe the {\it same physical state}. Conversely, only the
functions in phase space that are {\it invariant with respect to
gauge transformations} can describe physical quantities.

In order to eliminate this ambiguity and create a one-to-one mapping
between points in the phase space and physical states, we must
impose further constraints, known as {\it gauge conditions} or {\it
gauge-fixings}. The gauge-fixings can be implemented by arbitrary
functions of the canonical variables, except that they must
intersect each gauge orbit exactly once ({\it orbit conditions}) in
order to allow a well-posed definition of the {\it reduced phase
space}. The number of independent fixings must be equal to the
number of independent gauge variables (i.e., 8 in our case). The
canonical reduction follows a cascade procedure. Precisely, the
gauge-fixings to the {\it super-hamiltonian} and {\it
super-momentum} constraints come first (call it $\Gamma_4$): they
determine the 3-coordinate system and the {\it off-shell} shape of
$\Sigma_{\tau}$ (i.e., the {\it off-shell} convention for clock
synchronization); the requirement of their time constancy then
determines the gauge-fixings to the {\it primary} constraints: they
determine the {\it lapse} and {\it shift} functions. Finally, the
requirement of time constancy for these latter gauge-fixings
determines the Dirac multipliers. Therefore, the first level of
gauge-fixing gives rise to a {\it complete} gauge-fixing, say
$\Gamma_8$, and is sufficient to remove all the gauge arbitrariness.

\medskip

The $\Gamma_8$ procedure reduces the original 20-dimensional phase
space to a {\it copy} ${\Omega}_4$ of the {\it abstract reduced
phase-space} $\tilde{\Omega}_4$ having 4 degrees of freedom per
point (12 - 8 gauge-fixings). Abstractly, the reduced phase-space
$\tilde{\Omega}_4$ with its symplectic structure is defined by the
quotient of the constraint surface with respect to the 8-dimensional
group of gauge transformations and represents {\it the space of the
abstract gauge-invariant observables of GR: two configurational and
two momentum variables}. These observables carry the physical
content of the theory in that they represent the {\it autonomous
degrees of freedom of the gravitational field} (remember that at
this stage we are dealing with a pure gravitational field without
matter).

A $\Gamma_8$-dependent {\it copy} $\Omega_4$ of the {\it abstract}
$\tilde{\Omega}_4$ is realized in terms of the symplectic structure
(Dirac brackets) defined by the given gauge-fixings and
coordinatized by four DO [call such field observables $q^r$, $p_s$
(r,s = 1,2)]. The functional form of these DO ({\it concrete
realization of the gauge-invariant abstract observables in the given
complete gauge $\Gamma_8$}) in terms of the original canonical
variables depends upon the chosen gauge, so that such observables -
a priori - are neither tensors nor invariant under ${}_PDiff$. In
each gauge $\Gamma_8$, the original 8 gauge variables are now
uniquely determined functions of the DO. Yet, {\it off shell},
barring sophisticated mathematical complications, {\it any two
copies of $\Omega_4$ are diffeomorphic images of one another}.

\medskip

It is very important to understand qualitatively the geometric
meaning of the eight infinitesimal {\it off-shell} Hamiltonian gauge
transformations and thereby the geometric significance of the
related gauge-fixings. i) The transformations generated by the four
{\it primary} constraints modify the {\it lapse} and {\it shift}
functions which, in turn, determine both how densely the space-like
hyper-surfaces $\Sigma_{\tau}$ are distributed in space-time and the
appearance of {\it gravito-magnetism} on them; ii) the
transformations generated by the three {\it super-mo\-men\-tum}
constraints induce a transition on $\Sigma_{\tau}$ from one given
3-coordinate system to another; iii) the transformation generated by
the {\it super-hamiltonian} constraint induces a transition from one
{\it a-priori} given "form" of the 3+1 splitting of $M^4$ to another
(namely, from a given notion of distant simultaneity to another), by
operating deformations of the space-like hyper-surfaces in the
normal direction.

It should be stressed again that the {\it manifest effect} of the
gauge-fixings related to the above transformations emerges only
{\it at the end} of the canonical reduction and {\it after} the
solution of the Einstein-Hamilton equations has been worked out
(i.e. {\it on shell}). This happens because the role of the
gauge-fixings is essentially that of choosing the {\it functional
form} in which all the gauge variables depend upon the DO, i.e. -
physically - of fixing the {\it form} of the {\it inertial
potentials} of the associated {\it chrono-geometrically possible}
NIF. It must also be emphasized that this important physical
aspect is completely lost within the {\it abstract reduced phase
space} $\tilde{\Omega}_4$, which could play, nevertheless, another
important role (see Sections V and VI). It is only after the
initial conditions for the DO have been arbitrarily selected on a
Cauchy surface that the whole four-dimensional chrono-geometry of
the resulting {\it Einstein "universe"} is {\it dynamically}
determined, including the embedding functions $x^{\mu} =
z^{\mu}(\tau ,\vec \sigma)$ (i.e. the {\it on-shell} shape of
$\Sigma_{\tau}$), and therefore even the {\it dynamically
admissible} NIF within the set of {\it chrono-geometrically
possible} NIF. In particular, since the transformations generated
by the super-Hamiltonian modify the rules for the synchronization
of distant clocks, {\it all} the relativistic {\it conventions}
(including those for gravito-magnetism) associated to all NIF in a
given {\it Einstein "universe"}, turn out to be {\it
dynamically-determined, gauge-related options}\footnote{Unlike the
special relativistic case where the various possible conventions
are non-dynamical options.}.

\bigskip

\noindent Two important points must be emphasized.

i) In order to carry out the canonical reduction {\it explicitly},
before implementing the gauge-fixings we must perform a basic
canonical transformation at the {\it off-shell} level, the so-called
Shanmugadhasan transformation (Shanmugadhasan, 1973; Lusanna, 1993),
moving from the original canonical variables to a new basis
including the DO as a canonical subset\footnote{In practice, this
transformation is adapted to seven of the eight constraints
(Lusanna, 2001; DePietri, Lusanna, Martucci \& Russo, 2002): they
are replaced by seven of the new momenta whose conjugate
configuration variables are the gauge variables describing the {\it
lapse} and {\it shift} functions and the choice of the spatial
coordinates on the simultaneity surfaces. The new basis contains the
conformal factor (or the determinant) of the 3-metric, which is
determined by the super-Hamiltonian constraint (though as yet no
solution has been found for this equation, also called the
Lichnerowicz equation), and its conjugate momentum (the last gauge
variable whose variation describes the normal deformations of the
simultaneity surfaces, namely the changes in the clocks'
synchronization convention).}. It should be stressed here that it is
not known whether the Shanmugadhasan canonical transformation, and
therefore the GR observables, can be defined {\it globally} in C-K
space-times. In most of the spatially compact space-times this
cannot be done for topological reasons. A further problem is that in
field theory in general the status of the canonical transformations
is still heuristic. Therefore the only tool (viz. the Shanmugadhasan
transformation) we have for a systematic search of GR observables in
every type of space-time still lacks a rigorous definition. In
conclusion, {\it the mathematical basis of our analysis regarding
the objectivity of space-time structures is admittedly heuristic,
yet our arguments are certainly no more heuristic than the
overwhelming majority of the theoretical and/or philosophical claims
concerning every model of GR}.

\medskip

\noindent The Shanmugadhasan transformation is highly {\it
non-local} in the metric and curvature variables: although, at the
end, for any $\tau$, the DO are {\it fields} indexed by the
coordinate point $\sigma^a$, they are in fact {\it highly non-local
functionals of the metric and the extrinsic curvature over the whole
{\it off-shell} surface $\Sigma_{\tau}$}. We can write, {\it
symbolically} ($^3{\pi}^{cd}$ are the momenta conjugate to
${}^3g_{ab}$):

\begin{eqnarray}
q^r(\tau, \vec\sigma)
   &=& {{\mathcal{F}}_{[\Sigma_{\tau}]}}^r
  \bigr[(\tau, \vec\sigma)|\ {}^3g_{ab},
  {}^3{\pi}^{cd}\bigl]
 \nonumber \\
p_s(\tau, \vec\sigma)
  &=& {{\mathcal{G}}_{[\Sigma_{\tau}]}}_s
    \bigr[(\tau, \vec\sigma)|\ {}^3g_{ab},
    {}^3{\pi^{cd}}\bigl], \quad r,s = 1,2.
 \label{E4}
\end{eqnarray}

\medskip

ii) Since, as already mentioned, in {\it spatially compact}
space-times the original canonical Hamiltonian in terms of the ADM
variables is zero, the Dirac Hamiltonian happens to be written
solely in terms of the eight constraints and Lagrangian multipliers.
This means, however, that this Hamiltonian generates purely harmless
gauge transformations, so that {\it it cannot engender any real
temporal change}. Therefore, in spatially compact space-times, in a
completely fixed Hamiltonian gauge we have a vanishing Hamiltonian,
and the canonical DO are constants of the motion, i.e.
$\tau$-independent.

In such models of GR with {\it spatially compact} space-times
without boundary (nothing is known if there is a boundary), one must
re-introduce the {\it appearance of evolution} in a frozen picture.
Without deeply entering this debated topic (see the viewpoints of
Earman, 2002, 2003, Maudlin, 2002, Rovelli, 1991, 2002, as well as
the criticisms of Kuchar, 1992, 1993, and Unruh, 1991), we only add
a remark on the {\it problem of time}. In all of the globally
hyperbolic space-times (the only ones admitting a canonical
formulation), the mathematical time $\tau$, labeling the
simultaneity (and Cauchy) surfaces, must be related to some
empirical notion of time (astronomical ephemerides time, laboratory
clock,...). In a GR model with frozen picture there is no physical
Hamiltonian governing the evolution in $\tau$ \footnote{Unless,
following Kuchar (1993), one states that the super-Hamiltonian
constraint is not a generator of gauge transformations but an
effective Hamiltonian instead.} and consequently there exists the
problem of defining a local evolution in terms of a clock built with
GR observables (with a time monotonically increasing with $\tau$),
as well as the problem of parametrizing other GR observables in
terms of this clock\footnote{See the concept of evolving constants
of motion, and the partial and complete observables of Rovelli
(1991, 2002), as well as a lot of other different point of views.}.
\medskip

Our advantage point, however, is that, in the case of {\it spatially
non-compact} space-times of the C-K class, the generator of
$\tau$-temporal evolution is the {\it weak ADM energy}\footnote{The
ADM energy is a Noether constant of motion representing the {\it
total mass} of the instantaneous "3-universe", just one among the
ten asymptotic ADM Poincar{\`e} {\it charges} that, due to the
absence of super-translations, are the only asymptotic symmetries
existing in C-K space-times. Consequently, the Cauchy surfaces
$\Sigma_{\tau}$ must tend to space-like hyper-planes normal to the
ADM momentum at spatial infinity. This means that: (i) such
$\Sigma_{\tau}$'s are the {\it rest frame} of the instantaneous
"3-universe"; (ii) asymptotic inertial observers exist and should be
identified with the {\it fixed stars}, and (iii) an asymptotic
Minkowski metric is naturally defined. This {\it asymptotic
background} allows to avoid a split of the metric into a background
metric plus a perturbation in the weak field approximation. Due to
point ii) the C-K space-times provide a model of both the {\it solar
system} and our {\it galaxy} but, as yet, not a well-defined model
for cosmology. If gravity is switched off, the C-K space-times
collapse to Minkowski space-time and the ADM Poincar\`e charges
become the Poincar\`e special relativistic generators. {\it These
space-times provide, therefore, the natural model of GR for
incorporating particle physics which, in every formulation, is a
chapter of the theory of representations of the Poincar\`e group on
Minkowski space-time in inertial frames, the elementary particles
being identified by the mass and spin invariants}. If we change the
boundary conditions, allowing the existence of super-translations,
the asymptotic ADM Poincar\`e group is enlarged to the
infinite-dimensional asymptotic SPI group (Wald, 1984) and we lose
the possibility of defining the spin invariant. Note that in
spatially-compact space-times {\it with boundary} it could be
possible to define a boundary Poincar\`e group (lacking in the
absence of boundary), but we know of no result about this case. The
mathematical background of these results can be found in Lusanna,
2001; Lusanna \& Russo, 2002; DePietri, Lusanna, Martucci \& Russo,
2002; Agresti, DePietri, Lusanna \& Martucci, 2004, and references
therein.\hfill\break Let us add some further comments:\hfill\break
A) The fact that particle physics is defined in the spatially
non-compact Minkowski space-time implies that speaking of, e.g.,
nucleosynthesis in spatially compact cosmologies entails a {\it huge
extrapolation}.\hfill\break B) Classical string theories and
super-gravity theories include particles, but their quantization
requires the introduction of a background space-time for defining
the particle Fock space. The only well-developed form of
background-independent quantum gravity (loop quantum gravity),
obtained by quantizing either the connection or the loop
representation of GR, leads to a quantum formulation inequivalent to
Fock space, so that it is as yet not known how to incorporate
particle physics. We hope that our viewpoint, taking into account
the non-inertial aspects of GR, can be developed to the extent of
being able to reopen the program of canonical quantization of
gravity in a background independent way by {\it quantizing the DO
only}. See Alba \& Lusanna, (2005b), for a preliminary attempt to
define relativistic and non-relativistic quantum mechanics in
non-inertial frames in Galilei and Minkowski space-times,
respectively, in such a way that the gauge variables describing the
inertial effects (the {\it appearances}) remain
c-numbers.\hfill\break C) Finally, quantum field theory in
background curved space-times does not admit a useful particle
interpretation of its states due to the absence of the notion of
Fourier transform (no way of defining the sign of energy and the
usual Fock space). As a consequence, the particle notion is replaced
by the notion of detector and in this approach it is not clear how
to recover the results of particle physics needed for
astrophysics.}. Indeed, this quantity {\it does generate real
$\tau$-temporal changes of the canonical variables}, changes which
can subsequently be rephrased in terms of some empirical clock
monotonically increasing in $\tau$. It is important to stress that
since the density ${\cal E}_{ADM}(\tau, \vec{\sigma})$ of the weak
ADM energy $\int d^3 \sigma {\cal E}_{ADM} (\tau, \vec{\sigma})$
contains the potentials of the inertial forces explicitly, it is a
{\it gauge-dependent quantity}. This is nothing else than another
aspect of the gauge-dependence problem of the energy density in GR.

\medskip

Thus, the final Einstein-Dirac-Hamilton equations for the DO, in a
complete gauge, are

\begin{equation}
\dot{q}^r = \{q^r, H_{\mathrm{ADM}}\}^*, \quad \dot{p}_s = \{p_s,
H_{\mathrm{ADM}}\}^*, \quad r,s = 1,2,
 \label{E5}
\end{equation}

\noindent where $H_{\mathrm{ADM}}$ is intended to be the restriction
of the {\it weak ADM energy} to $\Omega_4$ and where the
$\{\cdot,\cdot\}^*$ are the Dirac brackets.
\medskip

In conclusion, within the Hamiltonian formulation, we found a class
of solutions in which - unlike what has been correctly argued by
Earman (Earman, 2002; Belot \& Earman, 1999, 2001) for
spatially-compact space-times - there is a {\it real, NIF-dependent,
temporal change}. But this of course also means that the {\it
frozen-time} picture, being model dependent, is not a {\it typical}
feature of GR.

On the other hand, it is not clear whether the formulation of a
cosmological model for GR is necessarily limited to spatially
compact space-times without boundary. As already said, our model is
suited for the solar system and the galaxy. It cannot be excluded,
however, that our asymptotic inertial observers (up to now
identifiable with the fixed stars) might be identified with the
preferred frame of the cosmic background radiation with our 4-metric
including some pre-asymptotic cosmological term.

\section{The intrinsic gauge and the dynamical individuation of point-events}

We know that only two of the ten components of the metric are
physically essential: it seems plausible then to suppose that only
this subset can act as an individuating field, and that the
remaining components play a different role.

Consider the following {\it four scalar invariant functionals} (the
eigenvalues of the Weyl tensor), written here in Petrov's compressed
notation (see, e.g., Kramer, Stephani, MacCallum \& Herlit, 1980):

\begin{eqnarray}
w_1 &=& \mathrm{Tr} \, (g W g W), \nonumber \\
w_2 &=& \mathrm{Tr} \, (g W \epsilon W), \nonumber \\
w_3 &=& \mathrm{Tr} \, (g W g W g W), \nonumber \\
w_4 &=& \mathrm{Tr} \, (g W g W \epsilon W),
 \label{E6}
\end{eqnarray}

\noindent where $g$ is the 4-metric, $W$ is the Weyl tensor, and
$\epsilon$ is the Levi--Civita totally antisymmetric tensor.

Bergmann and Komar (Komar, 1958; Bergmann \& Komar, 1960;
Bergmann, 1961, 1962) proposed a set of invariant {\it intrinsic
pseudo-coordinates} as four suitable functions of the
\par\noindent
$w_T$ \footnote{Modulo the equations of motion, the eigenvalues
$w_T$ are functionals of the 4-metric and its first derivatives.},

\begin{equation}
\hat{I}^{[A]} = \hat{I}^{[A]} \bigr[ w_T[g(x),\partial g(x)]
\bigl], \quad A = 0,1,2,3.
 \label{E7}
\end{equation}

\noindent Indeed, under the hypothesis of no space-time
symmetries, the $\hat{I}^{[A]}$ can be used to label the
point-events of space-time, at least locally. Since they are {\it
scalars}, the $\hat{I}^{[A]}$ are invariant under passive
diffeomorphisms (therefore they do not define a coordinate chart
in the usual sense, precisely as it happens with the {\it
radar-pseudo-coordinates}).

Clearly, our attempt to use intrinsic pseudo-coordinates to provide
a physical individuation of point-events would {\it prima facie}
fail in the presence of symmetries, when the $\hat{I}^{[A]}$ become
degenerate. This objection was originally raised by Norton (see
Norton, 1988, p.60) as a critique to manifold-plus-further-structure
(MPFS) substantivalism (see for instance Maudlin, 1988, 1990).
Several responses are possible. Firstly, although to this day all
the {\it known} exact solutions of Einstein's equations admit one or
more symmetries, these mathematical models are very idealized and
simplified; in a realistic situation (for instance, even with two
masses alone) space-time would be filled with the excitations of the
gravitational degrees of freedom, and would admit no symmetries at
all. Secondly, the parameters of the symmetry transformations can be
used as supplementary individuating fields, since, as noticed by
Stachel (1993), they also depend on the metric field, through its
isometries. Thirdly, and most importantly, in our analysis of the
physical individuation of points we are arguing a question of
principle, and therefore we must consider {\it generic} solutions of
Einstein's equations rather than the null-measure set of solutions
with symmetries.

\medskip
It turns out that the four Weyl scalar invariants can be
re-expressed in terms of the ADM variables, namely the lapse $N$ and
shift $N^a$ functions, the 3-metric ${}^3g_{ab}$ and its conjugate
canonical momentum (the extrinsic curvature ${}^3K_{a,b}$)
\footnote{Bergmann and Komar have shown that the four eigenvalues of
the {\it spatial part} of the Weyl tensor depend only upon the
3-metric and its conjugate momentum.}. Consequently the
$\hat{I}^{[A]}$ can be exploited to implement four gauge-fixing
constraints {\it involving a hyper-surface $\Sigma_{\tau}$ and its
embedding in $M^4$}. On the other hand, in a completely fixed gauge
$\Gamma_8$, the $\hat{I}^{[A]}$ become {\it gauge dependent}
functions of the DO of that gauge.
\medskip

Writing

\begin{equation}
\hat{I}^{[A]} [w_T(g, \partial g)] \equiv \hat{Z}^{[A]} [{\hat
w}_T({}^3g, {}^3\pi, N, N^a)], \quad A = 0,1,2,3;
 \label{E8}
\end{equation}

\noindent and selecting a {\it completely arbitrary,
radar-pseudo-coordinate system} $\sigma^A \equiv [\tau,\sigma^a]$
adapted to the $\Sigma_\tau$ surfaces, we apply the {\it intrinsic
gauge-fixing} defined by

\begin{equation}
\chi^A \equiv \sigma^A - \hat{Z}^{[A]} \bigl[ {\hat
w}_T[{}^3g(\sigma^B), {}^3\pi(\sigma^D), N(\sigma^E),
N^a(\sigma^F)] \bigr] \approx 0, \quad A, B, D, E, F = 0,1,2,3;
 \label{E9}
\end{equation}

\noindent to the {\it super-hamiltonian} (A = 0) and the {\it
super-mo\-men\-tum} (A = 1,2,3) constraints. This is a good
gauge-fixing provided that the functions $\hat{Z}^{[A]}$ are chosen
to satisfy the fundamental {\it orbit conditions}
$\{\hat{Z}^{[A]},\mathcal{H}_B\} \neq 0, \quad(A,B = 0,1,2,3)$,
which ensure the independence of the $\chi^A$ and {\it carry
information about the Lorentz signature}. Then the complete
$\Gamma_8$ {\it intrinsic gauge-fixing} procedure leads to the final
result

\begin{equation}
\sigma^A \equiv \tilde{Z}^{[A]} [ q^a(\sigma^B),  p_b(\sigma^D)|
\Gamma)], \quad A, B, D = 0,1,2,3;\quad a,b = 1,2;
 \label{E10}
\end{equation}

\noindent where the notation $|\Gamma)$ means the functional form
assumed in the chosen gauge $\Gamma_8$.

{\it On-shell} the last equation amounts to a {\it definition} of
the {\it radar-pseudo-coordinates} $\sigma^A$ as four {\it scalars}
providing a {\it physical individuation of any point--event, in
terms of the gravitational degrees of freedom $q^a$ and $p_b$}.
Therefore the scalars $\tilde{Z}^{[A]}$, {\it strongly identical} to
the radar pseudo-coordinates, define the enlarged Atlas ${\cal A}$
of $M^4$ referred to in footnote 22. In this way, each of the
point--events of space-time is endowed with its own {\it metrical
fingerprint} extracted from the tensor field, i.e. the value of the
four scalar functionals of the DO (exactly four!)\footnote{The fact
that there are just {\it four} independent invariants for the vacuum
gravitational field should not be regarded as a coincidence. On the
contrary, it is crucial for the purpose of point individuation and
for the gauge-fixing procedure we are proposing.}. The price that we
have paid for this achievement is that {\it we have necessarily
broken general covariance}! As already repeatedly stressed, every
choice of 4-coordinates for a point (every gauge-fixing, in the
Hamiltonian language), in any procedure whatsoever for solving
Einstein's equations, amounts to a breaking of general covariance,
by definition. On the other hand, the whole extent of general
covariance can be recovered by exploiting the gauge freedom. Our
construction does {\it not} depend on the selection of a set of
physically preferred intrinsic pseudo-coordinates, because by
modifying the functions $I^{[A]}$ we have the possibility of
implementing {\it any} (adapted) radar-coordinate system. Passive
diffeomorphism-invariance reappears in a different suit: we find
exactly the same functional freedom of ${}_PDiff\, M^4$ in the
functional freedom of the choice of the {\it pseudo-coordinates}
$Z^{[A]}$ (i.e. of the gauge-fixing). What matters here is that {\it
any} adapted radar-coordinatization of the manifold can be seen as
embodying the physical individuation of points, because it can be
implemented as the Komar--Bergmann {\it intrinsic
pseudo-coordinates} after we choose the correct $Z^{[A]}$ and select
the proper gauge.

\medskip

In conclusion, as soon as the Einstein-Dirac-Hamilton equations are
solved {\it in the chosen gauge $\Gamma_8$}, starting from given
initial values of the DO on a Cauchy hyper-surface
$\Sigma_{\tau_{0}}$, the evolution in $\tau$ throughout $M^4$ of the
DO themselves, whose dependence on space (and on parameter time) is
indexed by the chosen coordinates $\sigma^A$, yields the following
{\it dynamically-determined} effects: i) {\it reproduces} the
$\sigma^A$ as the Bergmann-Komar {\it intrinsic pseudo-coordinates};
ii) reconstructs space-time as a ({\it on-shell}) foliation of
$M^4$; iii) defines the associated {\it dynamically-admissible} NIF;
iv) determines a {\it simultaneity} and {\it gravito-magnetism
convention}.

Now, what happens if matter is present? Matter changes the Weyl
tensor through Einstein's equations and, in the new basis
constructed by the Shanmugadhasan transformation, contributes to the
separation of gauge variables from DO through the presence of its
own DO. In this case we have DO for both the gravitational field and
the matter fields, which satisfy {\it coupled
Einstein-Dirac-Hamilton equations}. Since the gravitational DO will
still provide the individuating fields for point-events according to
our procedure, {\it matter will come to influence the evolution of
the gravitational DO and thereby the physical individuation of
point-events and the dynamically-admissible NIF}. Of course, a basic
role of matter is the possibility of building apparatuses for the
measurement of the gravitational field, i.e. for an empirical
localization of point-events. As shown elsewhere (Pauri \&
Vallisneri, 2002; LPI and LPII), as a dynamical theory of
measurement is lacking, the epistemic circuit of GR can be
approximately closed via an {\it experimental three-step procedure}
that, starting from concrete radar measurements and using
test-objects, ends up in a complete and empirically coherent {\it
intrinsic individuating gauge fixing}, i.e. in an empirical
construction of a net of radar coordinates and in a measurement of
the metric in such coordinates.

\medskip

Finally, let us emphasize that, even in the case with matter, time
evolution is still ruled by the {\it weak ADM energy}. Therefore,
the temporal variation corresponds to a {\it real change} and not
merely to a harmless gauge transformation as in other models of GR.
The latter include, as already stressed in Section IV, the spatially
compact space-time without boundary. Note furthermore that, since
the DO of every completely fixed gauge in these spatially compact
models are $\tau$-independent, the gauge fixing with $A = 0$ in
(\ref{E10}) is inconsistent: it is therefore impossible to realize
the {\it time}-direction in terms of DO, and {\it the individuation
of point-events breaks down}. This is compatible with the
Wheeler-DeWitt interpretation according to which in such models we
have only a {\it local time evolution} (in the direction normal to
$\Sigma_{\tau}$) generated by the super-hamiltonian constraint (see
for instance Kuchar, 1993). It is seen that {\it our individuation
procedure fails in spatially compact models of GR on the same
grounds that prevent a real time evolution for them}. More
precisely, in such models it is possible to get at best a physical
individuation of the point-events belonging to the 3-space on a
fixed time slice, but not to space-time on the whole \footnote{
This, by the way, is just what happens in loop quantum gravity: one
starts with a fixed classical time slice (a Cauchy surface) as a
given 3-space and makes the quantization. This 3-space appears
explicitly in all the construction (see Nicolai, Peeters and
Zamaklar, 2005). Up to now there has been no accepted solution for
the issue of temporal evolution in spatially compact space-times
(the problem of the super-hamiltonian constraint) and, therefore, of
the spatiotemporal interpretation of the quantization.}.

\section{"The last remnant of physical objectivity of space-time": a final look}

The main results we have so far obtained are: i) a {\it
NIF-dependent temporal evolution} of the physical observables; ii)
the dis-solution of the Hole Argument; iii) a {\it NIF-dependent
physical individuation} of point-events in terms of the autonomous
degrees of freedom of the gravitational field (the {\it metrical
fingerprint} we were looking for); while results i) and iii) are
valid for the C-K class only, result ii) is valid for every
globally-hyperbolic space-time.

We want to scrutinize such results from the point of view of {\it
the issue of objectivity} of general relativistic space-time
structures.

Concerning the first result, we can only stress that the
NIF-dependence of the generator of temporal evolution is nothing
else than another manifestation of the endless problem of energy of
GR.

Concerning the searched for {\it explanation} of Leibniz
equivalence, our analysis of the correspondence between symmetries
of the Lagrangian configurational approach and those of the
Hamiltonian formulation has shown the following. Solutions of
Einstein's equations that, in the configurational approach, differ
within the Hole by elements of the subset ${}_ADiff^{'}\, M^4 $,
which correspond to mappings among gauge-equivalent Cauchy data,
belong to the {\it same 4-geometry}, i.e. the same equivalence class
generated by applying all {\it passive diffeomorphisms} to any of
the original 4-metrics: $ {}^4Geom  = {}^4Riem / {}_PDiff\, M^4 =
{}^4Riem / Q'$. In this case, as seen at the Hamiltonian level, they
are simply solutions differing by a {\it harmless Hamiltonian gauge
transformation on shell} and describing, therefore, the same
Einstein "universe". Furthermore, it is possible to engender these
allegedly different {\it models} of GR within the hole, by
appropriate choices of the initial gauge fixing (the functions
$\hat{Z}^{[A]}$). Since we know that the physical role of the
gauge-fixings is essentially that of choosing the functional form of
the inertial potentials in the NIF defined by the complete gauge
(the {\it epistemic} part of the game), the "differences" among the
solutions generated within the Hole by the allowed active
diffeomorphisms amount to the different {\it inertial appearances}
of the autonomous gravitational phenomena (the {\it ontic} part of
the game) in different NIFs.
\medskip

{\it In the end, this is what, physically, Leibniz equivalence
reduces to. This conclusion, together with the physical
individuation of point-events achieved by exploiting the intrinsic
gauge, make up our answer to Friedman's question. The extent of
intrinsic-ness of such an answer will be specified presently.}
\medskip

As already anticipated, our analysis contrasts with Stachel's
attitude towards the Hole Argument. Leaving aside Stachel's broad
perspective on the significance and the possibility of
generalizations of the Hole story (see Stachel \& Iftime, 2005), let
us confine ourselves to a few comments about Stachel's original
proposal for the physical individuation of points of $M^4$ by means
of a fully covariant exploitation of the Bergmann-Komar invariants
$\hat{I}^{[A]} \bigr[ w_T[g(x),\partial g(x)] \bigl], A = 0,1,2,3$.
First of all, remember again that the effect of the Hole Argument
reveals itself on {\it solutions} of Einstein's equations and that
the active diffeomorphisms that purportedly maintain the physical
identity of the points are, therefore, dynamical symmetries. Now,
how are we guaranteed that the functional dependence of the
covariant quantities $\hat{I}^{[A]} \bigr[ w_T[g(x),\partial g(x)]
\bigl]$ be {\it concretely characterized} as relating to actual
solutions of Einstein's equations ? Since in the actual case we know
that these quantities depend upon 4 DO and 8 gauge variables, we
have, hidden under general covariance, a gauge arbitrariness that
unavoidably transfers itself on the individuation procedure and {\it
leaves it undefined}. Indeed, speaking of general covariance in an
abstract way hides the necessity of getting rid of the above
arbitrariness by a gauge-fixing that, in turn, necessarily breaks
general covariance. In other words, a definite individuation entails
a concrete characterization of the {\it epistemic} part of the game,
which is precisely what we have done. The result is, in particular,
exactly what Stachel's suggestion was intended for, for our {\it
intrinsic gauge} shows that {\it active diffeomorphisms} of the
first kind (i.e. those belonging to $Q'$ in their passive
interpretation) do map individuations of point-events into {\it
physically} equivalent individuations. Indeed, since the {\it
on-shell} Hamiltonian gauge transformation connecting two different
gauges is the passive counterpart in $Q'$ of an {\it active}
diffeomorphisms $D_A \in {}_ADiff^{'}\, M^4$, it determines the {\it
drag-along coordinate transformation} ${\cal T}_{D_A}$ of Section II
connecting the 4-radar-coordinates of the two gauges, i.e. the {\it
dual view} of the active diffeomorphism. While the active
diffeomorphism {\it carries along} the identity of points by
assumption, its passive view {\it attributes different
physically-individuated radar-coordinates to the same (mathematical)
point}. It is seen, therefore, that for any point-event a given
individuation by means of DO is mapped into a physically-equivalent,
NIF-dependent, individuation.

It is worth stressing again that the main reason why we succeeded in
carrying out a concrete realization of Stachel's original suggestion
to its natural end lies in the possibility that the Hamiltonian
method offers of working {\it off-shell}. In fact, the 4-D active
diffeomorphisms, {\it qua} dynamical symmetries of Einstein's
equations, must act on solutions at every stage of the procedure and
fail to display the arbitrary {\it epistemic} part of the scalar
invariants. On the other hand, the Hamiltonian separation of the
gauge variables (characterizing the NIF and ruling the {\it
generalized inertial effects}) from the DO (characterizing
generalized tidal effects) {\it is an off-shell procedure} that
brings in the wanted {\it metrical fingerprint} by working
independently of the initial value problem. Once again, this
mechanism is a typical consequence of the special role played by
gauge variables in GR\footnote{According to a {\it main conjecture}
we have advanced elsewhere (see LPI \& LPII), a canonical basis of
{\it scalars} (coordinate-independent quantities), or at least a
Poisson algebra of them, should exist, making the above distinction
between DO and gauge variables {\it fully invariant}. An evaluation
of the degrees of freedom in connection with the Newman-Penrose
formalism for tetrad gravity (Stewart, 1993) tends to corroborate
the conjecture. In the Newman-Penrose formalism we can define ten
coordinate-independent quantities, namely the ten Weyl scalars. If
we add ten further scalars built using the extrinsic curvature, we
have a total of twenty scalars from which one should extract a
canonical basis replacing the 4-metric and its conjugate momenta.
Consequently, it should be possible to find {\it scalar} DO (the
{\it Bergmann observables}, see LPII) and some {\it scalar} gauge
variables (for instance a scalar version of the shift functions,
allowing a coordinate-independent description of gravito-magnetism).
In any case, the three gauge variables connected to the choice of
the 3-coordinates (or at least certain combinations of them) cannot
be made scalar, since they appear in those terms inside the ADM
energy density which describe the potentials of the intrinsically
coordinate-dependent inertial effects. The individuating functions
of (\ref{E8}) would depend on scalars only and the distinction
between DO and gauge observables would become {\it fully invariant}.
Yet, even in the case that the main conjecture might be proved, the
gauge-fixing procedure would always break general covariance and one
should not forget, furthermore, that the concept of {\it
radar-coordinates} contains a built-in {\it frame-dependence} (see
Section IV). Finally, and above all, the energy density ${\cal
E}_{ADM}(\tau, \vec{\sigma})$ would remain a NIF-dependent quantity
anyway.}.

Consider now the results achieved by exploiting the {\it intrinsic
gauge}. First of all, let us state that results iii) are derived on
the following assumptions only: a) the recourse to Hamiltonian
methods, which are necessary to keep the initial value problem under
control; b) the analysis of the $Q$ group of Bergmann \& Komar that
provides the unique way for connecting the Bergmann-Komar {\it
intrinsic pseudo-coordinates} to our {\it radar coordinates}. It
should, therefore, be stressed that the uniqueness of the
mathematical basis (the way in which the four scalar eigenvalues of
the Weyl tensor can be equated to four {\it scalar radar
pseudo-coordinates} by means of the {\it intrinsic gauge}) shows
that this methodology constitutes the only possible way of
disclosing the proper point-events ontology of the class of
space-times we are referring to. For given initial data of the DO
(identifying an Einstein's "universe"), any other kind of
gauge-fixing procedure would lead to gauge-equivalent solutions in
which the underlying point-events ontology simply {\it would not be
manifestly shown}. Therefore, there are no different formulations or
methodologies to compare that could affect the conclusions
(philosophical or not) to be drawn from the theory.

As to the physical individuation, our results are tantamount to
claiming that the {\it physical role of the gravitational field
without matter} is exactly that of individuating {\it physically}
the points of $M^4$  as point-events, by means of the four
independent phase-space degrees of freedom. As pointed out above,
the mathematical structure of the canonical transformation that
separates the DO from the gauge variables is such that the DO are
{\it highly non-local functionals of the metric and the extrinsic
curvature over the whole (off-shell) hyper-surface $\Sigma_\tau$}.
The same is clearly true for the {\it intrinsic pseudo-coordinates}
[see Eq.(\ref{E8})].

This said, we can even state that the existence of physical {\it
point-events} in our models of general relativity appears to be
synonymous with the existence of the DO for the gravitational field.
We advance accordingly the {\it ontological} claim that - physically
- {\it Einstein's vacuum space-time in our models is literally
identifiable with the autonomous degrees of freedom of such a
structural field}, while the specific (NIF-dependent) functional
form of the {\it intrinsic pseudo-coordinates} associates such
coordinates to the points of $M^4$. The autonomous gravitational
degrees of freedom are - so to speak - {\it fully absorbed in the
individuation of point-events}. On the other hand, when matter is
present, the individuation methodology maintains its validity and
shows how matter comes to influence the physical individuation of
point-events.

At this point, looking back at our results on their whole but with a
special attention to iii), we should make a clear assessment of the
{\it degree of physical objectivity} of our individuation procedure
{\it vis a' vis} the radical statement made by Einstein in the
passage quoted in the Introduction. What matters, of course, is the
NIF-dependence of the {\it physical identity} of point-events we
have achieved for a given Einstein's "universe". Clearly, a {\it
really different} physical individuation is obtained starting with
different initial conditions for the {\it Dirac observables} (i.e.
for a different {\it "universe"}).

\medskip

Now, can the freedom of choice among the dynamically-possible NIF be
equated with "taking away from space and time the last remnant of
physical objectivity" as Einstein claimed ? We do believe that the
answer is certainly "no". On the other hand, however, we should
acknowledge that what we have gained does not seem to be a kind of
objectivity in the usual sense, since the values of the DO which
individuate the point-events are NIF-dependent. It is only in the
{\it abstract reduced phase space} $\tilde{\Omega}_4$, defined in
Section IV by taking the quotient with respect to all the gauge
inertial effects, that {\it abstract} DO live: they are the {\it
"strictly intrinsic" qualifiers} of the generalized tidal effects
(i.e. of the proper degrees of freedom of the gravitational field),
which are then concretely realized by the ordinary Dirac observables
in each NIF. For any given Einstein's "universe" with its topology,
the abstract DO in $\tilde{\Omega}_4$ are locally functions of the
points $x$ of an abstract mathematical manifold $\tilde{M}_4$ that
is the equivalence class of all our {\it concrete realizations} of
space-time, each one equipped with its gauge-dependent individuation
of points, NIF and inertial forces. The point over which such fields
reside could be called {\it intrinsic} to the extent that they are
no longer NIF-dependent, and synthesize the essential properties of
all the appearances shown by the gauges. Admittedly, the global
existence of $\tilde{\Omega}_4$ over $\tilde{M}_4$ is subjected to a
huge set of mathematical hypotheses which we will not take into
account here. Locally, however, the Dirac fields certainly exist and
we could introduce a coordinate system defined by their values as
intrinsic individuating system for the given "universe". Given the
abstract nature of the NIF-independent DO, these considerations
possess a purely mathematical value. At any rate, we take them as
meaningful enough to justify the affix "point" to our notion of
"point-structuralism"\footnote{Recall the following passage by
Bergmann and Komar: "[...] in general relativity the identity of a
world point is not preserved under the theory's widest invariance
group. This assertion forms the basis for the conjecture that some
physical theory of the future may teach us how to dispense with
world points as the ultimate constituents of space-time altogether."
(Bergmann \& Komar, 1972, p.27). Now the {\it abstract reduced phase
space} $\tilde{\Omega}_4$ would be just the germ of such a theory.
The theory would be an {\it abstract and highly non-local theory} of
classical gravitation that, transparency aside, would be stripped of
all the {\it epistemic machinery} (the gauge freedom) which is
indispensable for both an empirical access to the theory and the
reconstruction of the {\it local field}   $g_{\mu \nu}(x)$. In other
words - inversely seen - the {\it gauge} structure contributes to
the {\it re-construction} of the spatiotemporal local and continuum
representation. We see that even in the context of classical
gravitational theory, the spatiotemporal {\it continuum} plays the
role of an {\it epistemic precondition} of our sensible experience
of macroscopic objects, playing a role which is not too dissimilar
from that enacted by Minkowski {\it micro-space-time} in the local
relativistic quantum field theory (see Pauri, 2000).}.

The important question, however, is another one, namely, what kind
of would-be fully objective spatiotemporal structures of our models
of GR should the discovered NIF-dependence of the {\it point-events
individuation} be compared to ? Here, again, the fascination of
general covariance must be put to the test, and again we will find
that the problem has to do with an excessively self-confident
utilization of the configurational (and fully general covariant)
geometric interpretation of a complex mathematical notion like that
of Cauchy-surface. This point was dealt with in great detail in
LPII, Section 3. Here we shall limit ourselves to exploiting the
notion of {\it Bergmann observable} (BO) as an 'acid test' for any
off-hand attribution of "true objectivity", or NIF-independence, or
"unique predictability", to spatiotemporal structures in GR.
Briefly, a BO is a configurational quantity defined in $M^4$ which
is both coordinate-independent (i.e. it is a scalar field or an
invariant under ${}_PDiff\ $) and also {\it "uniquely predictable
from initial data"}. The essential point is that - under the current
general covariance wisdom - quantities which are "invariant" under
the passive diffeomorphisms are often confidently, but wrongly,
taken to be also "uniquely predictable from the initial data". In
LPII,3 we considered, in particular, the example of the
four-dimensional scalar curvature $R(p)$, calculated at any point
$p$ of $M^4$ lying in the "future" of a Cauchy-surface. This
quantity, taken to be a BO, has been considered by Earman (see
Earman, 2002) to the effect of showing that the observables in GR
cannot undergo {\it any kind of change at all}, let alone {\it
temporal change}. Now, the problem is exactly the same raised above
for the Bergmann-Komar scalars in Stachel's perspective: how could
we be sure by definition that a scalar field - {\it qua} fully
covariant entity - when rewritten in terms of ADM variables, does
not contain arbitrary gauge elements that can jeopardize the
conclusion to be drawn from its seemingly symmetric simplicity ? For
this is exactly what happens in this case: as a matter of fact,
$R(p)$ turns out to be a {\it gauge-dependent} (or NIF-dependent)
quantity and therefore {\it not predictable}; in conclusion {\it it
is not a BO}! This lack of predictability, however, cannot be
perceived within the configurational and fully covariant approach of
$M^4$: the shallow geometric interpretation of the Cauchy surface
{\it fails the 'acid test'}. The same NIF-dependence characterizes,
for instance, quantities like the following {\it off-shell scalars}
with respect to ${}_PDiff\, M^4$: the bilinears
${}^4R_{\mu\nu\rho\sigma}\, {}^4R^{\mu\nu\rho\sigma}$, ${}^4R
_{\mu\nu\rho\sigma}\, \epsilon^{\mu\nu\alpha\beta}\,
{}^4R_{\alpha\beta}{} ^{\rho\sigma}$ and - as already said - the
{\it four eigenvalues of the Weyl tensor} exploited in Section V.
What is more important is that the same {\it does hold}, in
particular, for the {\it one-way velocity of light}, for {\it the
line element} $ds^{2}$ and, therefore, for the very {\it causal
structure of space-time}.

This technical de-tour highlights the important fact that, as soon
as one leaves the rarefied atmosphere of full general covariance and
soils his hands with the dirty facts of the empirical front of GR,
i.e. - theoretically -  with the {\it epistemic component} of the
inertio-gravitational field, one realizes that {\it all} of the
fundamental features of space-time structure are - in our language -
NIF-dependent. While the local equivalence principle dissolves the
absolute structures of the special theory of relativity, the {\it
global consequences} of the (local) equivalence principle are
precisely the pervasive non-inertials factors that show themselves
as NIF-dependence. It seems therefore reasonable that we do not
expect for {\it the physical individuation of point-events}, in our
models, a degree of objectivity greater than that of all the other
physically relevant structures in GR. To conclude, we will state
that all of these structures are - unvoidably - {\it weakly, or
NIF-objective}.

\medskip

We would like to surmise that the disclosure of the physical meaning
of Leibniz equivalence renders even more glaring the ontological
diversity of the gravitational field with respect to all other
fields, even beyond its prominent causal role. It seems
substantially difficult to reconcile the nature of the gravitational
field with the standard approach of theories based on a {\it
background space-time} (to wit, string theory and perturbative
quantum gravity in general). Any attempt at linearizing such
theories unavoidably leads to looking at gravity from the
perspective of a spin-2 theory in which the graviton stands at the
same ontological level as other quanta. In the standard approach of
background-dependent theories of gravity, photons, gluons and
gravitons all live on the stage on an equal footing. From the point
of view set forth in this paper, however, {\it non-linear gravitons}
are at the same time both the stage and the actors within the causal
play of photons, gluons, and other {\it material characters} such as
electrons and quarks.

\medskip

Note finally that the individuating relation (\ref{E10}) is a
numerical identity that has an {\it in-built non-commutative
structure}, deriving from the Dirac--Poisson structure hidden in its
right-hand side. The individuation procedure transfers, as it were,
the non-commutative Poisson-Dirac structure of the DO onto the
individuated point-events, even though the coordinates on the l.h.s.
of the identity are c-number quantities. One could guess that such a
feature might deserve some attention in view of quantization, for
instance by maintaining that the identity, interpreted as a relation
connecting mean values, could still play some role at the quantum
level.

\bigskip

\section{Concluding remarks: An instantiation of
structural realism as "point-structuralism"}

We conclude by spending a few words about the implications of our
results for some issues surrounding the recent debate on scientific
structural realism, as well as for the traditional debate on the
absolutist/relationist dichotomy.

It is well-known that the term {\it scientific realism} has been
interpreted in a number of different ways within the literature on
philosophy of science, in connection with the progressive
sophistication of our understanding of scientific knowledge. Such
ways concern, e.g., realism about {\it observable or unobservable
entities}, and realism about {\it theories}. A further ramification
of meanings has been introduced more recently by the so-called {\it
structural realism} (the only attainable reality are relations
between (unobservable) objects), which originated a division between
the so-called {\it epistemic} structural realists (entity realism is
unwarranted) and the {\it ontic} structural realists (the relations
exhaust what exists) (see Simon, 2003).

From the logical point of view, we can assume that the concept of
{\it structure} refers to a (stable or not) set of {\it relations}
among a set of some kind of {\it constituents} that are put in
relations (the {\it relata}). The specification expressed by the
notion of {\it structural realism} introduces {\it some kind} of
ontological distinction between the role of the relations and that
of the constituents. At least two main exemplary possibilities
present themselves as obvious: (i) there are relations in which the
constituents are (ontologically) primary and the relation secondary;
(ii) there are relations in which the relation is (ontologically)
primary while the constituents are secondary, and this even without
any prejudice against the ultimate ontological consistency of the
constituents. In the case of {\it physical} entities, following
Stachel (see Stachel, 2005) one could cautiously recover in this
connection the traditional distinction between {\it essential} and
{\it non-essential} properties ({\it accidents}) in order to
characterize the {\it degree of (ontological) primacy} of the
relations versus the {\it relata} and vice versa (and this
independently of any metaphysical flavor possibly connected to the
above distinctions). For example, one could say that in the extreme
case (i) only {\it accidental} properties of the constituents can
depend upon the relational structure, while in the extreme case (ii)
at least one {\it essential} property of the constituents depends
upon the relational structure (saying that {\it all} the essential
properties of the relata depend upon the relation would be
tantamount to claiming that there exist only relations without
constituents, as the {\it ontic structural realist} has it).

A further complication is connected to the nature of the structure
we are considering. For while at the logical level (leaving aside
the deep philosophical issue concerning the relationships between
mathematical structures and substances) the concept of {\it
mathematical structure} (e.g. a system of differential equations, or
even the bare mathematical manifold of point which provides the
first layer of our {\it representations} of the real space-time )
can be taken to be sufficiently clear for our purposes, the
definition of {\it physical structure} immediately raises
existential philosophical problems. For example, we believe that it
is very difficult to define a {\it physical structure} without
bringing in its {\it constituents}, and thereby granting them {\it
some kind of existence} and defending some sort of {\it entity
realism}. Analogously, we believe that it is very difficult to
defend {\it structural realism} without also endorsing a {\it theory
realism} of some sort. However, both theses are not universally
shared.

Having said this, let us come back to the results we obtained in the
previous sections. The analysis based on our {\it intrinsic gauge}
has disclosed a remarkable and rich local structure of the
general-relativistic space-time for the considered models of GR. In
correspondence to every {\it intrinsic gauge} (\ref{E10}) we
achieved a gauge-related {\it physical individuation of
point-events} in terms of the DO of that gauge, i.e. in terms of the
{\it ontic} part of the gravitational field, as represented in the
clothes furnished by the NIF. Such individuation is characterized by
a highly {\it non-local} functional dependence of the DO upon the
values of the metric and the extrinsic curvature over the whole
({\it off-shell}) space-like hyper-surface $\Sigma_{\tau}$ of
distant simultaneity. Since the extrinsic curvature has to do with
the embedding of the simultaneity hyper-surface in $M^4$, the DO do
{\it involve geometrical elements external to the hyper-surface}
itself. In fact, the temporal gauge (fixed by the scalar $Z^{[0]}$)
in the identity (\ref{E10}) refers to a continuous interval of
hyper-surfaces, and the gauge-fixing identity itself is {\it
intrinsically four-dimensional}. We have, therefore, an
instantiation of {\it metrical holism} which, though local in the
temporal dimension and characterized by a {\it dynamic
stratification in 3-hypersurfaces}, is {\it four-dimensional}.
Admittedly, the distinction between {\it ontic} and {\it epistemic}
parts, as well as the {\it form} of the space-like surfaces of
distant simultaneity, are NIF-dependent.

Thus we have discovered that ontologically the identity of
point-events is {\it conferred} upon them by a complex relational
structure in which they are holistically enmeshed. This relational
structure includes all the elements of the {\it complete gauge
fixing $\Gamma_8$} summarized by a NIF, and supported by a definite
solution of Einstein's equations throughout $M^4$, corresponding to
given initial values for the DO in that gauge (a definite Einstein
"universe"). We propose to define such physical identity of
point-events as {\it weakly-objective} or {\it NIF-objective}, with
the important notice, however, that this weak degree of objectivity
is the maximum that can be attained for {\it all} of the relevant
spatiotemporal structures in a formulation in which the initial
value problem for the C-K models of GR is well-posed. Although the
{\it holistic structure} appears to be {\it ontologically prior} to
its {\it constituents} (the point-events) as to their physical
identity, we cannot agree with Cao's assertion (see Cao, 2003,
p.111) that the constituents, as mere {\it place-holders}, derive
their meaning or {\it even their existence} from their function and
place in the structure. Indeed, at any level of GR, the empirical
level above all, one cannot avoid quantifying over points, and we
have just attributed a physical meaning\footnote{Even operationally,
in principle (see Pauri \& Vallisneri, 2002; LPI and LPII, already
quoted in Section V).} to the radar-coordinate indexing of such
point-events. Such an indexing makes the latter as ontologically
equivalent to the existence of the gravitational field in vacuum as
an extended entity, since the autonomous degrees of freedom of the
gravitational field are - so to speak - {\it fully absorbed in the
individuation of point-events}. From this point of view, our dressed
point-events are not under-determined by empirical evidence. Quite
in general, we cannot see how a place-holder can have any
ontological function in an evolving network of relationships without
possessing at least {\it some kind} of properties. For our results
do in fact confer a sort of causal power on the
gravitationally-dressed points.

Furthermore, as already said, we could even dare a little more
concerning the identity of points, since some kind of {\it abstract
intrinsic individuality} survives beneath the variety of
descriptions displayed by all the gauge-related NIF and common to
all these {\it appearances}. This kind of {\it intrinsic identity} -
in the vein suggested by Bergmann and Komar - is just furnished by
the {\it abstract Dirac fields} residing within the phase-space
$\tilde{\Omega}_4$, which is nothing else but a quotient with
respect to all of the concrete realizations and appearances of the
NIF. Accepting this suggestion for the sake of argument we would be
led to a peculiar space-time {\it structure} in which the
relation/relata correspondence does not fit with any of the extreme
cases listed above, for one could assert that while the abstract
{\it essential} properties belong to the constituents as seen in
$\tilde{\Omega}_4$ (so that abstract point-events in $\tilde{M^4}$
would be like {\it natural kinds}), the totality of the physically
concrete {\it accidents} are displayed by means of the holistic
relational structure. This is the reason why we propose calling this
peculiar kind of space-time structuralism {\it point-structuralism}.
Admittedly, it is important not to be misled into thinking that this
abstract intrinsic-ness has a direct physical meaning.

Summarizing, this view holds that space-time point-events (the {\it
relata}) do exist and we continue to quantify over them; however,
their properties can be viewed both as {\it extrinsic} and {\it
relational}, being conferred on them in a holistic way by the whole
structure of the metric field and the extrinsic curvature on a
simultaneity hyper-surface, and, at the same time - at an abstract
level only - as {\it intrinsic}, being coincident with the
autonomous degrees of freedom of the gravitational field represented
by the abstract NIF-independent Dirac fields in $\tilde{\Omega}_4$.
In this way - although point-events cannot be viewed as {\it genuine
individuals} - both the metric field and point-events maintain their
{\it own manner of existence}, so that the structural texture of
space-time in our models does not force us to abandon an {\it entity
realist} stance about both the metric field and its points. We must,
therefore, deny the thesis according to which metrical relations can
exist without their constituents (the point-events).

\medskip

Concerning the traditional debate on the dichotomy
substantivalism/relationism, we believe that our analysis may indeed
offer a {\it tertium-quid} solution to the debate.

First of all, let us recall that, in remarkable diversity with
respect to the traditional historical presentation of Newton's
absolutism {\it vis \'a vis} Leibniz's relationism, Newton had a
much deeper understanding of the nature of space and time. In two
well-known passages of {\it De Gravitatione}, Newton expounds what
could be defined an original {\it proto-structuralist view} of
space-time (see also Torretti, 1987, and DiSalle, 1994). He writes
(our {\it emphasis}):

\begin{quotation}
{\footnotesize \noindent Perhaps now it is maybe expected that I
should define extension as substance or accident or else nothing
at all. But by no means, for it has {\it its own manner of
existence} which fits neither substance nor accidents [\ldots] The
parts of space derive their character from their positions, so
that if any two could change their positions, they would change
their character at the same time and each would be converted
numerically into the other {\it qua} individuals.  The parts of
duration and space are only understood to be the same as they
really are because of their mutual order and positions ({\it
propter solum ordinem et positiones inter se}); nor do they have
any other {\it principle of individuation} besides this order and
position which consequently cannot be altered. (Hall \& Hall,
1962, p.99, p.103.)}
\end{quotation}

\noindent On the other hand, in his relationist arguments, Leibniz
could exploit the conjunction of the Principle of Sufficient Reason
and the Principle of the Identity of Indiscernibles because
Newtonian space was {\it uniform}, as the following passage lucidly
explains (our {\it emphasis}):

\begin{quotation}

{\footnotesize \noindent Space being {\it uniform}, there can be
neither any {\it external} nor {\it internal reason}, by which to
distinguish its parts, and to make any choice between them. For, any
external reason to discern between them, can only be grounded upon
some internal one. Otherwise we should discern what is
indiscernible, or choose without discerning. (Alexander, 1956,
p.39).}
\end{quotation}

\noindent Clearly, if the parts of space were real, Leibniz
Principles would be violated. Therefore, for Leibniz, space is not
real. The upshot, however, is that space (space-time) in general
relativity, far from being {\it uniform} may possess, as we have
seen, a {\it rich structure}. This is just the reason why - in our
sense - it is {\it real}, and why Leibniz equivalence called upon
for general relativity happens to hide the very nature of
space-time, instead of disclosing it.

We claim that our results lead to a new kind of {\it structuralist}
conception of space-time. Such structuralism is not only richer than
that of Newton, as one could expect because of the dynamical
structure of Einstein space-time, but richer in an even deeper
sense. Not only the independent degrees of freedom of the metric
field are able to characterize the "mutual order and positions" of
points {\it dynamically}, since each point-event "is" - so to speak
- the "values" of the autonomous degrees of freedom of the
gravitational field; their capacity is even stronger, since {\it
such mutual order is altered by the presence of matter}.

This {\it new structuralist} conception turns out to include
elements common to the tradition of both {\it substantivalism} and
{\it relationism}. Although the metric field does {\it not} embody
the traditional notion of {\it substance}, it is taken to represent
a genuine and primitive element of physical reality and its
definition is a necessary condition in order to be able even to {\it
speak} of space-time. In this sense {\it exists} and plays a role
for the individuation of point-events by means of its {\it
structure}. On the other hand, our {\it point-structuralism} does
not support even the standard relationist view. In fact, the
holistic relationism we defend does not reduce the whole of
spatiotemporal relations to physical relations (i.e. it is not
eliminativist), nor does it entails that space-time does not exist
as such, being reducible to physical relations. Our dressed
point-events are "individuals" in a peculiar sense: they exist as
autonomous constituents, but one cannot claim that their properties
do not depend on the properties of others. Not only relations, but
also their carriers do exist, even if they do bring intrinsic
properties in a very special sense.

Let us finally consider what John Earman wrote in his 1989 book:
\begin{quotation}
\footnotesize \noindent The absolute-relational contrast is far from
being a dichotomy. A possible, third alternative, which I shall call
the property-view of space-time, would take something from both
camps: it would agree with the relationist in rejecting a
substantival substratum for events while joining with the absolutist
in recognizing monadic properties of spatiotemporal locations.
(Earman, 1989, p.14).
\end{quotation}

\noindent

Clearly, {\it stricto sensu}, we cannot say that our view of
space-time is a {\it property-view} in Earman's sense. All
quantities definable with respect to our gravitationally-dressed
point-events are not irreducible monadic spatiotemporal properties.
For they are reducible on the grounds that the {\it physical}
identity of our point-events is NIF-relational. Still, due to the
underlying abstract structure of autonomous gravitational degrees of
freedom represented by the abstract NIF-independent Dirac fields in
$\tilde{\Omega}_4$, we feel allowed to talk of a {\it weak-property
view} of space-time.

We acknowledge that the validity of our results is restricted to the
class of models of GR we worked with. Yet, we were interested in
exemplifying a question of principle, so that we can claim that
there is a class of models of GR embodying both a {\it real notion
of NIF-dependent temporal change}, a {\it NIF-dependent physical
individuation of points} and a new {\it structuralist and holistic
view of space-time}.

\medskip

\section{Acknowledgments}

We thank our friend Mauro Dorato for many stimulating discussions

\vfill\eject

\section{References}

Alba,D. and Lusanna, L.(2003). Simultaneity, radar 4-coordinates
and the 3+1 point of view about accelerated observers in special
relativity, (http://lanl.arxiv.org/abs/gr-qc/0311058).
\medskip

Alba,D. and Lusanna,L. (2005a). Generalized radar 4-coordinates and
equal-time Cauchy surfaces for arbitrary accelerated observers,
(http://lanl.arxiv.org/abs/gr-qc/0501090). \medskip

Alba,D. and Lusanna,L. (2005b). Quantum mechanics in non-inertial
frames with a multi-temporal quantization scheme: I) relativistic
particles (http://lanl.arxiv.org/abs/hep-th/0502060); II)
non-relativistic particles
(http://lanl.arxiv.org/abs/hep-th/0504060). \medskip

Agresti,J., De Pietri,R., Lusanna,L. and Martucci,L. (2004).
Hamiltonian linearization of the rest-frame instant form of tetrad
gravity in a completely fixed 3-orthogonal gauge: a radiation gauge
for background-independent gravitational waves in a post-Minkowskian
Einstein space-time, {\it General Relativity and Gravitation} {\bf
36}, 1055-1134; (http://lanl.arxiv.org/abs/gr-qc/0302084).\medskip .

Alexander,H (ed.),(1956).{\it The Leibniz-Clarke correspondence},
fourth paper. Manchester: Manchester University Press. \medskip

Arnowitt,R., Deser,S., and Misner,C.W. (1962). The dynamics of
general relativity, in L. Witten (ed.), \emph{Gravitation: an
introduction to current research}, (pp. 227--265). NewYork: Wiley.
\medskip

Bartels,A. (1994). What is spacetime if not a substance? Conclusions
from the new Leibnizian argument, in U. Mayer and H.-J. Schmidt
(eds.), \emph{Semantical aspects of spacetime theories},
(pp.41--51). Mannheim: B.I. Wissenshaftverlag.
\medskip

Belot,G. and Earman,J. (1999). From metaphysics to physics, in
J.Butterfield and C.Pagonis, (eds.), \emph{From physics to
philosophy}, (pp. 167--186). Cambridge: Cambridge University
Press.\medskip

Belot,G. and Earman,J.(2001). Pre-Socratic quantum gravity, in
C.Callender (ed.), {\it Physics meets philosophy at the Planck
scale. Contemporary theories in quantum gravity}. (pp. 213--255).
Cambridge: Cambridge University Press.\medskip

Bergmann, P.G. and Komar, A. (1960). Poisson brackets between
locally defined observables in general relativity, {\it Physical
Review Letters} {\bf 4}, 432-433.\medskip

Bergmann, P.G. (1961). Observables in general relativity, {\it
Review of Modern Physics} {\bf 33}, 510-514\medskip

Bergmann, P.G. (1962). The general theory of relativity, in S.Flugge
(ed.), {\it Handbuch der Physik}, Vol. IV, {\it Principles of
electrodynamics and relativity}, (pp. 247-272). Berlin:
Springer.\medskip

Bergmann,P.G. and Komar, A. (1972). The coordinate group symmetries
of general relativity, {\it International Journal of Theoretical
Physics} {\bf 5}, 15-28.\medskip

Butterfield,J. (1984). Relationism and possible worlds,
\emph{British Journal for the Philosophy of Science} \textbf{35},
1--13.\medskip

Butterfield,J.(1987) Substantivalism and determinism,
\emph{International Studies in the Philosophy of Science}
\textbf{2}, 10--31.\medskip

Butterfield,J.(1988). Albert Einstein meets David Lewis, in A.
Fine and J. Leplin (eds.), \emph{PSA 1988}, \textbf{2}, (pp.
56--64).\medskip

Butterfield,J.(1989). The hole truth, \emph{British Journal for the
Philosophy of Science} \textbf{40}, 1--28.\medskip

Cao,T.(2003), Can we dissolve physical entities into mathematical
structure ?, {\it Synth\'ese}, \textbf{136}, 51-71\medskip

Christodoulou, D., and Klainerman, S. (1993). {\it The global
nonlinear stability of the Minkowski space}. Princeton: Princeton
University Press.\medskip

De Pietri,R., Lusanna,L., Martucci,L. and Russo,S. (2002). Dirac's
observables for the rest-frame instant form of tetrad gravity in a
completely fixed 3-orthogonal gauge, {\it General Relativity and
Gravitation} {\bf 34}, 877-1033;
(http://lanl.arxiv.org/abs/gr-qc/0105084).\medskip

DeWitt,B. (1967) Quantum theory of gravity, I) the canonical theory,
{\it Physical Review} {\bf 160}, 1113-1148. II) the manifestly
covariant theory, {\bf 162}, 1195-1239.\medskip

DiSalle,R. (1994) On dynamics, indiscernibility, and spacetime
ontology", {\it British Journal for the Philosophy of Science,
vol.45,}, pp.265-287.
\medskip

Dorato,M (2000) Substantivalism, relationism, and structural
spacetime realism, {\it Foundations of Physics}, {\bf 30},
pp.1605-1628.
\medskip

Dorato,M and Pauri,M. (2006) Holism and structuralism in classical
and quantum general relativity, Pittsburgh-archive, ID code 1606,
in (2006). D.Rickles, S.French \& J.T.Saatsi (eds.), {\it The
structural foundations of quantum gravity}, forthcoming (November
2, 2006) Oxford: Clarendon Press.\medskip

Earman,J. and Norton,J. (1987). What price spacetime
substantivalism? The hole story, \emph{British Journal for the
Philosophy of Science} \textbf{38}, 515--525.\medskip

Earman,J. (1989). \emph{World enough and space-time}. Cambridge,
Mass.: The Mit Press.\medskip

Earman,J. (2002). Thoroughly modern McTaggart or what McTaggart
would have said if he had read the general theory of relativity,
\emph{Philosophers' Imprint} \textbf{2}, No.3,
(http://www.philosophersimprint.org/002003/).\medskip

Earman, J. (2003). Ode to the constrained Hamiltonian formalism, in
K.Brading, and E.Castellani, (eds.), {\it Symmetries in physics:
philosophical reflections}. Cambridge: Cambridge University
Press.\medskip

Einstein,A. (1914). Die formale Grundlage der allgemeinen
Relativit\"atstheorie, in \emph{Preuss. Akad. der Wiss. Sitz.},
(pp. 1030--1085).\medskip

Einstein,A. (1916). Die Grundlage der allgemeinen
Relativit\"atstheorie, \emph{Annalen der Physik} \textbf{49},
769--822; (1952) translation by W. Perrett and G. B. Jeffrey, The
foundation of the general theory of relativity, in \emph{The
principle of relativity}, (pp. 117--118). New York: Dover.
\medskip

Friedman, M.,(1984), {\it Roberto Torretti, relativity and
geometry}, Critical Review, \emph{No\^us} \textbf{18}, 653--664.
\medskip

Friedman, M.,(2002), Geometry as a branch of physics: background
and context for Einstein's "Geometry and experience", in {\it
Reading natural philosophy: essays in the history and philosophy
of science and mathematics to honor Howard Stein on his $70^{th}$
birthday}, D.Malament (ed.). Chicago: Open Court.
\medskip

Friedrich, H. and Rendall, A. (2000). The Cauchy problem for
Einstein equations, in B.G.Schmidt (ed.), {\it Einstein's field
equations and their physical interpretation}. Berlin: Springer;
(http://lanl.arxiv.org/abs/gr-qc/0002074).\medskip

Hall, A.R. and Hall, M.B., (eds.), (1962). {\it De gravitatione et
aequipondio fluidorum, unpublished scientific papers of Isaac
Newton. A selection from the Portsmouth collection in the university
library}. Canbridge: Cambridge University Press.
\medskip

Hawking, W.S. and Horowitz, G.T. (1996), The gravitational
Hamiltonian, action, entropy and surface terms, {\it Classical and
Quantum Gravity} {\bf 13}, 1487-1498.\medskip

Hilbert,D. (1917) Die Grundlagen der Physik. (Zweite Mitteilung),
\emph{Nachrichten von der K\"oniglichen Gesellschaft der
Wissenschaften zu G\"ottingen, Mathematisch-physikalische Klasse},
(pp. 53--76).\medskip

Howard, D. and Norton, J. (1993), Out of the labyrinth? Einstein,
Hertz, and G\"ottingen answer to the hole argument, in \emph{The
attraction of gravitation: new studies in the history of general
relativity}, Einstein studies, Vol. 5, J. Earman, M. Jansen, J.
Norton, eds., Boston: Birkh\"auser, pp. 30--62.
\medskip

Komar,A. (1958). Construction of a complete set of independent
observables in the general theory of relativity, {\it Physical
Review} {\bf 111}, 1182-1187.\medskip

Kramer,D., Stephani,H., MacCallum,M. and Herlit,E. (1980). {\it
Exact solutions of Einstein's field equations}, Cambridge monographs
on mathematical physics, 6, Cambridge: Cambridge University Press.
\medskip

Kuchar, K. (1992). Time and interpretations of quantum gravity, in
{\it Proceedings of the 4th Canadian conference on general
relativity and relativistic astrophysics}, (pp. 211-314). Singapore:
World Scientific.\medskip

Kuchar, K. (1993). Canonical quantum gravity, 13th Conference on
general relativity and gravitation (GR-13), Cordoba, Argentina, 29
Jun - 4 Jul 1992. In G.Kunstatter, D.E.Vincent, and J.G.Williams
(eds.), {\it Cordoba 1992, general relativity and gravitation},
(pp.119-150);(http://lanl.arxiv.org/abs/gr-qc/9304012).\medskip

Lusanna, L. (1993). The Shanmugadhasan canonical transformation,
function groups and the second Noether theorem, {\it International
Journal of Modern Physics} {\bf A8}, 4193-4233.\medskip

Lusanna, L. (2001). The rest-frame instant form of metric gravity,
{\it General Relativity and Gravitation} {\bf 33}, 1579-1696;
(http://lanl.arxiv.org/abs/gr-qc/0101048).\medskip

Lusanna,L. and Russo,S. (2002) A new parametrization for tetrad
gravity, {\it General Relativity and Gravitation} {\bf 34}, 189-242;
(http://lanl.arxiv.org/abs/gr-qc/0102074).\medskip

Lusanna, L. and Pauri, M. (2006-I) The physical role of
gravitational and gauge degrees of freedom in general relativity. I:
dynamical synchronization and generalized inertial effects, {\it
General Relativity and Gravitation}, ({\bf 38})2, pp. 187-227;
(http://lanl.arxiv.org/abs/gr-qc/0403081).\medskip

Lusanna,L. and Pauri,M. (2006-II). The physical role of
gravitational and gauge degrees of freedom in general relativity.
II: Dirac versus Bergmann observables and the objectivity of
space-time, {\it General Relativity and Gravitation}, ({\bf 38})2,
pp. 229-267; (http://lanl.arxiv.org/abs/gr-qc/0407007).\medskip

Maudlin,T. (1988). The essence of space-time, in \emph{PSA 1988},
\textbf{2}, (pp. 82--91).\medskip

Maudlin,T. (1990). Substances and spacetimes: what Aristotle would
have said to Einstein, \emph{Studies in the History and Philosophy
of Science} \textbf{21}, 531--61.\medskip

Maudlin, T. (2002). Thoroughly muddled McTaggart: or how to abuse
gauge freedom to generate metaphysical monstrosities. {\it
Philosophers' Imprint}, {\bf 2}, No. 4.\medskip

Nicolai, H., Peeters, K. and Zamaklar, M. (2005). Loop quantum
gravity: an outside view, hep-th/0501114]. \medskip

Norton,J. (1987). Einstein, the hole argument and the reality of
space, in J.Forge (ed.), {\it Measurement, realism and objectivity},
Reidel, Dordrecht.\medskip

Norton,J.(1988). The hole argument, in PSA (1988), vol.2, pp.
56-64.\medskip

Norton,J.(1992). The physical content of general covariance, in J.
Eisenstaedt, and A. Kox (eds.), {\it Studies in the history of
general relativity}, Einstein studies, vol. 3, (pp. 281--315).
Boston: Birkh\"auser.\medskip

Norton,J.(1993). General covariance and the foundations of general
relativity: eight decades of dispute, {\it Rep.Prog.Phys.} {\bf 56},
791-858.\medskip

Pauri,M. (1996) Realt\`a e oggettivit\`a, in F.Minazzi (ed.)
\emph{L'Oggettivit\`a nella conoscenza scientifica}, (pp.79--112).
Brescia: Franco Angeli.

\medskip

Pauri,M. (2000). Leibniz, Kant and the {\it quantum}. In E.Agazzi
and M.Pauri (eds.), {\it The reality of the unobservable}, Boston
studies in the philosophy of science, n.215, Dordrecht: Kluwer
Academic Publishers, (pp.270-272).

\medskip

Pauri,M. and M.Vallisneri,M. (2002). Ephemeral point-events: is
there a last remnant of physical objectivity?, Essay in honor of the
70th birthday of R.Torretti, {\it Dialogos} {\bf 79}, 263-303;
(http://lanl.arxiv.org/abs/gr-qc/0203014).\medskip

Rendall, A. (1998). Local and global existence theorems for the
Einstein equations, {\it Online Journal Living Reviews in
Relativity} {\bf 1}, n. 4; {\it ibid}. (2000) {\bf 3}, n. 1;
(http://lanl.arxiv.org/abs/gr-qc/0001008).\medskip

Rovelli, C. (1991). What is observable in classical and quantum
gravity. {\it Classical and Quantum Gravity}, {\bf 8},
297-316.\medskip

Rovelli, C. (2002). Partial observables. {\it Physical Review},
{\bf D, 65}, 124013, 1-8;
(http://lanl.arxiv.org/abs/gr-qc/0110035).\medskip

Rynasiewicz,R. (1994). The lessons of the hole argument,
\emph{British Journal for the Philosophy of Science} \textbf{45},
407--436.\medskip

Rynasiewicz,R. (1996). Absolute versus relational space-time: an
outmoded debate?, \emph{Journal of Philosophy} \textbf{43},
279--306.\medskip

Schlick,M. (1917) {\it Raum und Zeit in der gegenw\"artigen Physik}.
Berlin: Springer; third edition 1920, fourth edition 1922;
translated by H.Brose from third edition, {\it Space and time in
contemporary physics} (Oxford: Oxford University Press, 1920;
expanded to include changes in the fourth edition by P.Heath, in
H.Mulder and B. van de Velde-Schlick, (eds.), {\it Moritz Schlick:
philosophical papers}, vol.1. Dordrecht: Reidel, (1978),
pp.207-69.\medskip

Shanmugadhasan,S. (1973). Canonical formalism for degenerate
Lagrangians, {\it Journal of Mathematical Physics} {\bf 14},
677-687.\medskip

Simons,J. (ed).(2003). {\it Symposium on structural realism and
quantum field theory}, {\it Synth\'ese} {\bf 136}. N.1.\medskip

Stachel,J. (1980). Einstein's search for general covariance,
1912--1915. Ninth international conference on general relativity and
gravitation, Jena.\medskip

Stachel,J. (1993) The meaning of general covariance -- the hole
story, in J. Earman, I. Janis, G. J. Massey and N. Rescher,
(eds.), \emph{Philosophical problems of the internal and external
worlds, essays on the philosophy of Adolf Gr\"unbaum},
(pp.129--160). Pittsburgh: University of Pittsburgh Press.
\medskip

Stachel,J. (2005) Structure, individuality and quantum gravity,
gr-qc/0507078 v2. \medskip

Stachel,J. and Iftime,M. (2005) Fibered manifolds, natural bundles,
structured sets, G-sets and all that: the hole story from space-time
to elementary particles, gr-qc/0505138.\medskip

Stewart,J. (1993). {\it Advanced general relativity}, Cambridge:
Cambridge University Press.\medskip

Torretti,R. (1987). {\it Relativity and geometry}, New York: Dover,
pp.167-168.\medskip

Unruh,B. (1991). No time and quantum gravity. In R.Mann \& P.Wesson
(eds.), {\it Gravitation: A Banf Summer Institute}. Singapore: World
Scientific (pp.260-275).\medskip

Wald,R.M. (1984) {\it General relativity}. Chicago: University of
Chicago Press.\medskip

Weyl,H.,(1946). Groups, Klein's Erlangen program. Quantities, ch.I,
sec.4 of {\it The Classical groups, their invariants and
representations}, 2nd ed., (pp.13-23). Princeton: Princeton
University Press.

\end{document}